\documentclass[12pt,preprint]{emulateapj}
\usepackage{natbib}

\def\lap{\lower.5ex\hbox{$\; \buildrel < \over \sim \;$}}
\def\gap{\lower.5ex\hbox{$\; \buildrel > \over \sim \;$}}
\def\Msun{\hbox{M$_{\odot}$}}
\def\Lsun{\hbox{L$_{\odot}$}}
\def\micron{$\mu$m}
\def\kms{km s$^{-1}$}
\def\Msolaryr{\Msun yr$^{-1}$}

\begin{document}
\shorttitle{Infrared Luminosity Function and Star Formation Rate of the Bullet Cluster}
\shortauthors{Chung et al.}

\title{Star Formation in the Bullet Cluster I: The Infrared Luminosity Function and Star Formation Rate\footnote{This paper includes data gathered with the 6.5 meter Magellan Telescopes located at Las Campanas Observatory, Chile.}\footnote{Based on observations made with ESO Telescopes at La Silla under program ID 072.A-0511.}}

\author{Sun Mi Chung\altaffilmark{1}, Anthony H. Gonzalez\altaffilmark{1}, Douglas Clowe\altaffilmark{2,3}, Maxim Markevitch\altaffilmark{4}, Dennis Zaritsky\altaffilmark{5}}

\altaffiltext{1}{Department of Astronomy, University of Florida, Gainesville, FL 32611-2055; schung@astro.ufl.edu}
\altaffiltext{2}{Department of Physics and Astronomy, Ohio University, 251B Clippinger Lab, Athens, OH 45701}
\altaffiltext{3}{Alfred P. Sloan Fellow}
\altaffiltext{4}{Harvard-Smithsonian Center for Astrophysics, 60 Garden Street, Cambridge, MA 02138}
\altaffiltext{5}{Steward Observatory, University of Arizona, 933 North Cherry Avenue, Tucson, AZ 85721}

\begin{abstract}

The Bullet Cluster is a massive galaxy cluster at $z=0.297$ undergoing a major supersonic (Mach 3) merger event.  Using { \it Spitzer} 24\micron\ images, IRAC data, optical imaging, and optical spectroscopy, we present the global star formation rate (SFR) of this unique cluster.  Using a 90\% spectroscopically complete sample of 37 non-AGN MIPS confirmed cluster members out to $R<1.7$ Mpc, and the \citet{rieke2009} relation to convert from 24\micron\ flux to SFR, we calculate an integrated obscured SFR of 267 \Msolaryr\ and a specific star formation rate of 28 \Msolaryr\ per $10^{14}$\Msun.   The cluster mass normalized integrated SFR of the Bullet Cluster is among the highest in a sample of eight other clusters and cluster mergers from the literature.  Five LIRGs and one ULIRG contribute 30\% and 40\% of the total SFR of the cluster, respectively.   To investigate the origin of the elevated specific SFR, we compare the infrared luminosity function (IR LF) of the Bullet Cluster to those of Coma (evolved to z=0.297) and CL1358+62.  The Bullet Cluster IR LF exhibits an excess of sources compared to the IR LFs of the other massive clusters.  A Schechter function fit of the Bullet Cluster IR LF yields $L^{*}=44.68\pm0.11$ ergs s$^{-1}$, which is $\sim$0.25 and 0.35 dex brighter than $L^{*}$ of evolved Coma and CL1358+62, respectively.  The elevated IR LF of the Bullet Cluster relative to other clusters can be explained if we attribute the ``excess'' star-forming IR galaxies to a population associated with the infalling group that have not yet been transformed into quiescent galaxies.   In this case, the timescale required for quenching star formation in the cluster environment must be longer than the timescale since the group's accretion -- a few hundred million years.  We suggest that ``strangulation'' is likely to be an important process in the evolution of star formation in clusters.

\end{abstract}

\keywords{galaxies: clusters: individual (1E0657-56, The Bullet Cluster) -- galaxies: evolution}

\section{Introduction}

The fraction of star-forming galaxies in galaxy clusters is known to be suppressed relative to what is typically found in the less dense environments of galaxy groups and the general field population.  The overwhelming observational evidence \citep{dressler1980,gomez2003} suggests that star formation must be quenched at some point during hierarchical mass assembly.  Galaxy transformations are thought to occur predominantly in one of two broad categories -- prior to cluster assembly within the galaxy group environment, also known as galaxy pre-processing \citep{zabludoff1998,kodama2001}, or within the cluster environment itself.  While there is evidence in favor of both scenarios \cite[e.g.][]{fujita2004,cortese2006,koyama2008,berrier2009}, in the end the role of the cluster environment on the star formation history of its galaxies is unclear.  Even less clear is the impact of a major cluster merger on galaxy evolution.

\citet{owen2005,miller2003,ferrari2005}  found evidence for triggered star formation in cluster mergers based on an enhanced fraction of star-forming radio galaxies and a preferential distribution of emission line galaxies between merging subclumps.  In contrast, \citet{poggianti2004} found that the post-starburst population of dwarf galaxies in the Coma cluster lie near the edges of two merging substructures, suggesting that the merger quenches star formation, though these could be relics of starbursts induced during an earlier phase of the merger \citep{mahajan2010}.

There has also been much effort placed in understanding the specific physical mechanisms responsible for transforming galaxy properties such as morphology and star formation.  Although there is some observational and/or theoretical support for various physical processes considered to occur in the cluster environment, such as ram pressure \citep{gunn1972}, strangulation \citep{kawata2008}, and galaxy harassment \citep{moore1996}, it is still unknown which of these play a dominant role in transforming a star-forming galaxy into a quiescent one.

This is the second in a series of papers to examine the star formation properties of 1E0657-56, also known as the Bullet Cluster. The Bullet Cluster is a galaxy cluster at $z=0.297$ undergoing a major merger event, with a collision between the main cluster and subcluster occuring close to the plane of the sky with $i<8^{\circ}$ \citep{markevitch2004}.  A well-defined bow shock front has been confirmed by \citet{markevitch2002}, and is propagating through the X-ray gas in the subcluster region at a velocity of 4740$\pm630$ \kms~(Mach number $M=3.0\pm0.4$) \citep{markevitch2006}.  The subcluster itself lags behind the shock front due to a wind from the main cluster gas and travels at 2700 \kms\ relative to the main cluster \citep{springel2007}.  At this velocity, the time since core passage is $\sim$0.25 Gyr ago.

With our extensive multi-wavelength dataset and optical spectroscopy, we study star formation out to nearly the virial radius of the cluster {\it ($R_{200}=2.2$ Mpc)}.  In this series of papers, we will examine various aspects of star formation in the Bullet Cluster to understand how a recent major merger event affects the star formation of cluster galaxies.  We also consider the importance of different physical mechanisms that may dominate galaxy evolution in the diverse environments present in the Bullet Cluster.  Star formation in the core of the Bullet Cluster has been studied by \citet{chung2009} who have found that the strong ram pressure exerted by supersonic gas from the cluster merger does not have a significant impact on recent star formation in the cluster galaxies.

In this paper we use the {\it Spitzer} Multiband Imaging Photometer (MIPS) in the 24\micron\ band \citep{rieke2004}, to compare the global star formation rate of the Bullet Cluster, as traced by the dusty star-forming galaxies, to those of other merging and non-merging clusters from the literature.  We also present the infrared luminosity function of the Bullet Cluster as a means of investigating the nature of the infrared luminous galaxy population in this massive major cluster merger.

\section{Observations and Data Reduction}

The data used in this paper include  3.6\micron\ to 8\micron, and 24\micron\ imaging from the {\it Spitzer Space Telescope} and optical imaging from the 2.2m ESO telescope.   The {\it Spitzer} data cover a radius of $R\sim 1.7$ Mpc from the center of the Bullet Cluster field ($\alpha=6^{h}58^{m}26.^{s}0248, \delta=-55^{\circ}56^{'},49.^{''}3185$), while the optical data extend to $R\sim4.5$ Mpc.  We also use optical spectroscopy obtained over several years with the Inamori Magellan Areal Camera and Spectrograph (IMACS) to determine cluster membership for the mid-infrared {\it Spitzer} sources based on our catalog of 857 known redshifts in the Bullet Cluster field.  Throughout the paper we assume a cosmology of $H_{0}=71$ km/s/Mpc, $\Omega_{M}=0.27, \Lambda=0.73$.

\subsection{Spitzer MIPS}

We obtained observations of the Bullet Cluster on 30 November 2007, in the 24 $\mu$m band of the {\it Spitzer} Multiband Imaging Photometer \cite[MIPS;][]{rieke2004} in photometry mode. We observed in two cycles, with 14 frames at 30 seconds per frame for a total integration time of $\sim$940 s per pixel in the main exposed region.  Using a raster map in which frames are dithered along the scan and cross-scan directions, with half-array offsets, we achieved a total spatial coverage of 
$\sim12.^{'}6\times12.^{'}6$, which corresponds to a radius of 1.7 Mpc from the center of the Bullet Cluster.  Individual frames were combined into a final mosaic with the {\it Spitzer} Science Center (SSC) data analysis tool MOsaicker and Point source EXtractor \cite[MOPEX;][]{makovoz2005}.  MOPEX also includes a tool for photometry called APEX, which performs point-response function (PRF) fitting of point sources.  Using MOPEX and APEX, we obtained a catalog of 418 sources detected at a minimum of 5$\sigma$ above the background noise.

We determine the completeness of the MIPS catalog via artificial star tests.  A bright, unsaturated star in the mosaic is used as the point source function (PSF).  This ``PSF'' is then duplicated and scaled to create artificial stars that cover a range of fluxes.  Artificial stars are added to the original mosaic, then APEX is used for source detection.  This process is repeated until a smooth completeness function is obtained.  The 50\% and 80\% completeness limits are 134$\mu$Jy and 310$\mu$Jy, with average corresponding signal-to-noise (S/N) ratios of $\sim$11 and $\sim$23, respectively. 

\subsection{Spitzer IRAC}

We obtained data from the {\it Spitzer} InfraRed Array Camera \citep[IRAC;][]{fazio2004}  in all four bands -- 3.6$\mu$m, 4.5$\mu$m, 5.8$\mu$m, and 8.0$\mu$m on 14 November 2007 in full array readout mode with a small scale, cycling dither pattern.  With eight offset pointings, the total coverage is $15^{\prime}\times15^{\prime}$ in the four IRAC bands.

We created IRAC mosaics from individual frames using MOPEX, with source detection and photometry done using Source Extractor \citep{bertin1996}.  We compute galaxy colors using aperture photometry within a 6 pixel ($3.^{''}6$) diameter, which is sufficiently large for robust photometry while avoiding contamination from potential nearby neighbors in a crowded field.  No aperture corrections were applied because they are  minimal when considering colors ($\sim$0.05 magnitudes).

\subsection{WFI Imaging}

We obtained R, V, and B-band imaging using the Wide Field Imager (WFI) on the ESO/MPG 2.2m telescope at La Silla, Chile.  The data, collected in service-observing mode in January 2004, have total exposure times in the three filters of 14100s, 6580s, and 5640s, respectively, and cover an area of $34^{\prime}\times 34^{\prime}$ centered on the Bullet Cluster field.  Individual images have 470s exposure times, and were dithered between images to cover the chip gaps.  We reduced the images following the prescription of \citet{clowe2001}, with the addition of a photometric correction based on the scattered light present in flat fields \citep{koch2004}.  Photometric zero-points were determined using standard star fields from \citet{landolt1992}.  We aligned all images to a common coordinate system and construct a photometric catalog using SExtractor on the R-band image for object detection and R-band photometry, and the two-image mode to obtain photometry on the B and V-band images.

\subsection{IMACS Spectroscopy}

We have a total of 1122 optical spectra in the Bullet Cluster field in the years 2005, 2006, and 2009.  In 2005 and 2006, red sequence galaxies were the primary targets, while in 2009 we targetted MIPS sources and blue-cloud galaxies beyond the MIPS FOV.  We used the Inamori Magellan Areal Camera and Spectrograph (IMACS) on the 6.5m Magellan Baade telescope with the f/2 camera.   All spectra cover a wavelength range of 4000-9000 \AA\, with a dispersion of 2\AA\ per pixel and were reduced with {\it COSMOS}, the standard data reduction package for IMACS spectra.

\begin{figure}
\epsscale{1}
\plotone{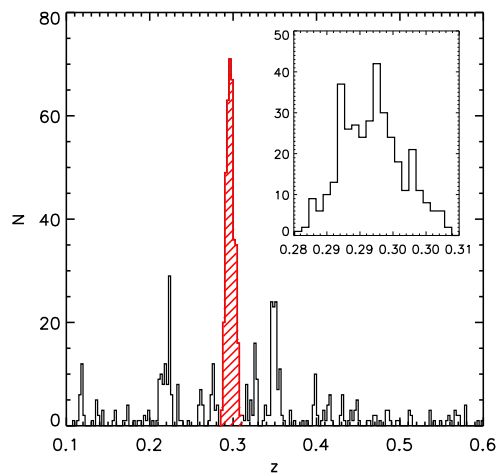}
\caption{Spectroscopic redshift distribution with shading to indicate cluster member region.  Inset shows zoomed-in view of cluster members as defined by the caustic analysis of the infall region.}
\label{fig:veldisp}
\end{figure}

To obtain redshifts, we use the IRAF task {\it xcsao} to cross-correlate spectra with template spectra of four galaxy types -- giant elliptical, spiral, E+A, and emission line galaxy.  Of the 1122 spectra, we  recover redshifts for 857 galaxies, of which 362 are designated cluster members based on a caustic analysis of the cluster infall region (Gonzalez et al. in prep), similar to the technique employed by \citet{diaferio2005}.  To our catalog of 362 confirmed members, we add another 44 members from \citet{barrena2002}. Figure~\ref{fig:veldisp} shows the distribution of spectroscopic redshifts, with the cluster members highlighted in red.  In addition to the Bullet Cluster, we find two prominent redshift peaks at $z\sim0.21$ and $z\sim0.35$, containing roughly 90 members each.  The {\it Herschel Space Observatory} far-infrared and sub-mm properties of the background group are studied in relation to the Bullet Cluster by Rawle et al (2010, in press). 

\section{Analysis}
\subsection{MIPS candidate members}

To quantify the total star formation rate of the Bullet Cluster as traced by 24\micron\ luminosity, we first identify the MIPS sources that are likely to be cluster members.  We use WFI and IRAC images to exclude galaxies whose colors are indicative of either background galaxies or active galactic nuclei (AGN).  We then refined our sample by spectroscopically targeting MIPS sources that are candidate cluster members.

\begin{figure*}
\epsscale{1}
\plottwo{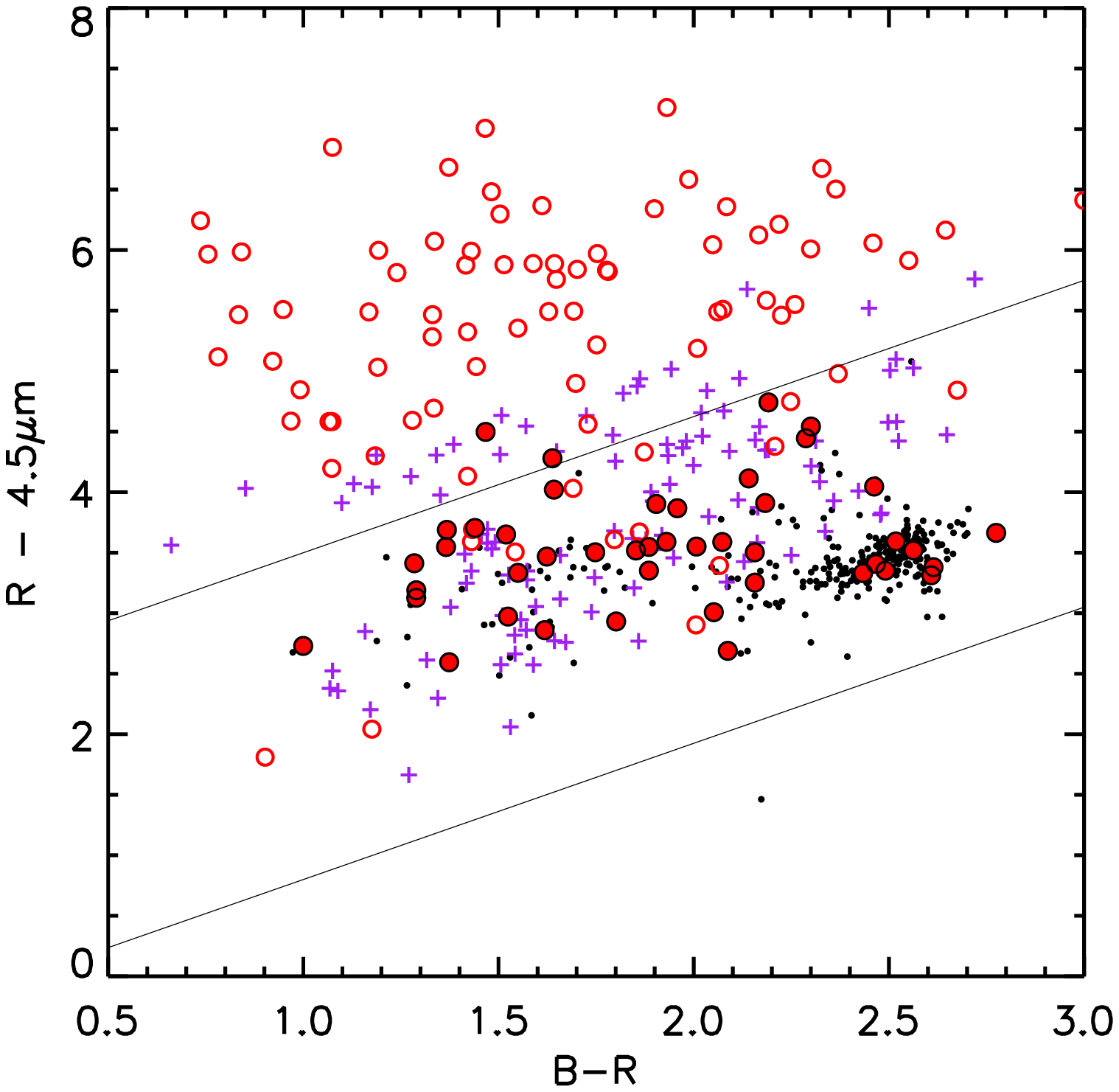}{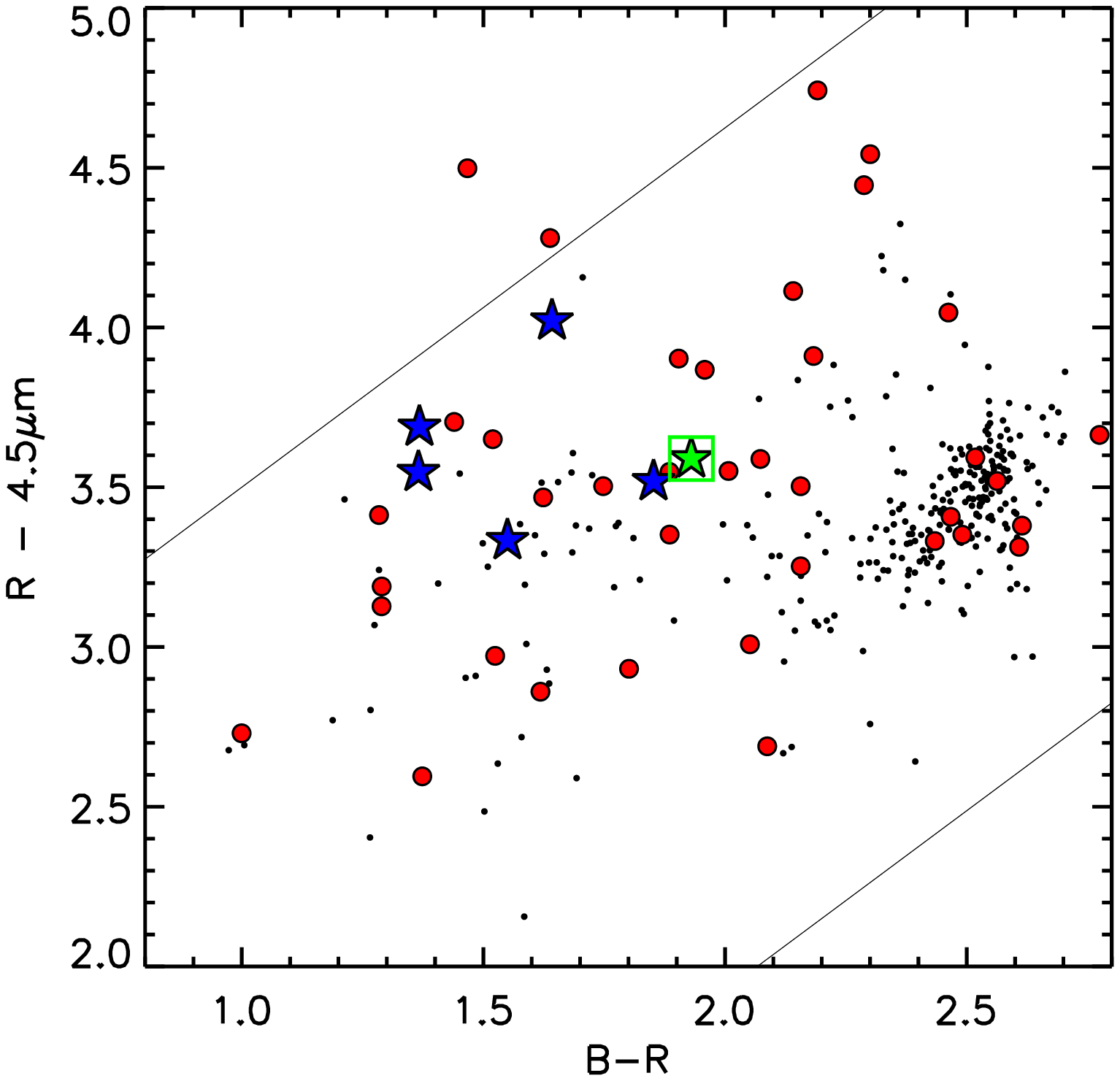}
\caption{The R-[4.5] and B-R colors of spectroscopically confirmed cluster members (dots), cluster members with MIPS emission (solid circle), cluster candidates with MIPS emission (open circle), and known interlopers with MIPS emission (cross).   The spectroscopically confirmed cluster galaxies, particularly those with MIPS emission, form a tight sequence in this color-color space.  To minimize contamination from background MIPS sources, we include galaxies only within the diagonal boundaries indicated.  Galaxies with $B-R>3$ are also excluded since these are optically redder than the red sequence.  The right panel shows the R-[4.5] and B-R colors of confirmed cluster members only.  Five confirmed cluster LIRGs are shown as blue stars, and one ULIRG as a green star with a square border.}
\label{fig:select}
\end{figure*}

We start with an initial MIPS catalog of 418 sources down to a flux of $43\mu$Jy. We match the WFI (BVR) and IRAC catalogs and then cross-match with the MIPS catalog. Figure~\ref{fig:select} shows the R-[4.5] versus B-R colors for all MIPS sources with optical and IRAC counterparts. Spectroscopically confirmed members from the 2005, 2006, and first part of 2009 IMACS campaigns demonstrated that the star-forming cluster members form a tight diagonal locus in the R-[4.5] vs B-R space. We define our cluster candidate sample as galaxies within the two solid diagonal lines shown in Figure~\ref{fig:select}, and blueward of B-R=3.

\begin{figure*}
\epsscale{1}
\plottwo{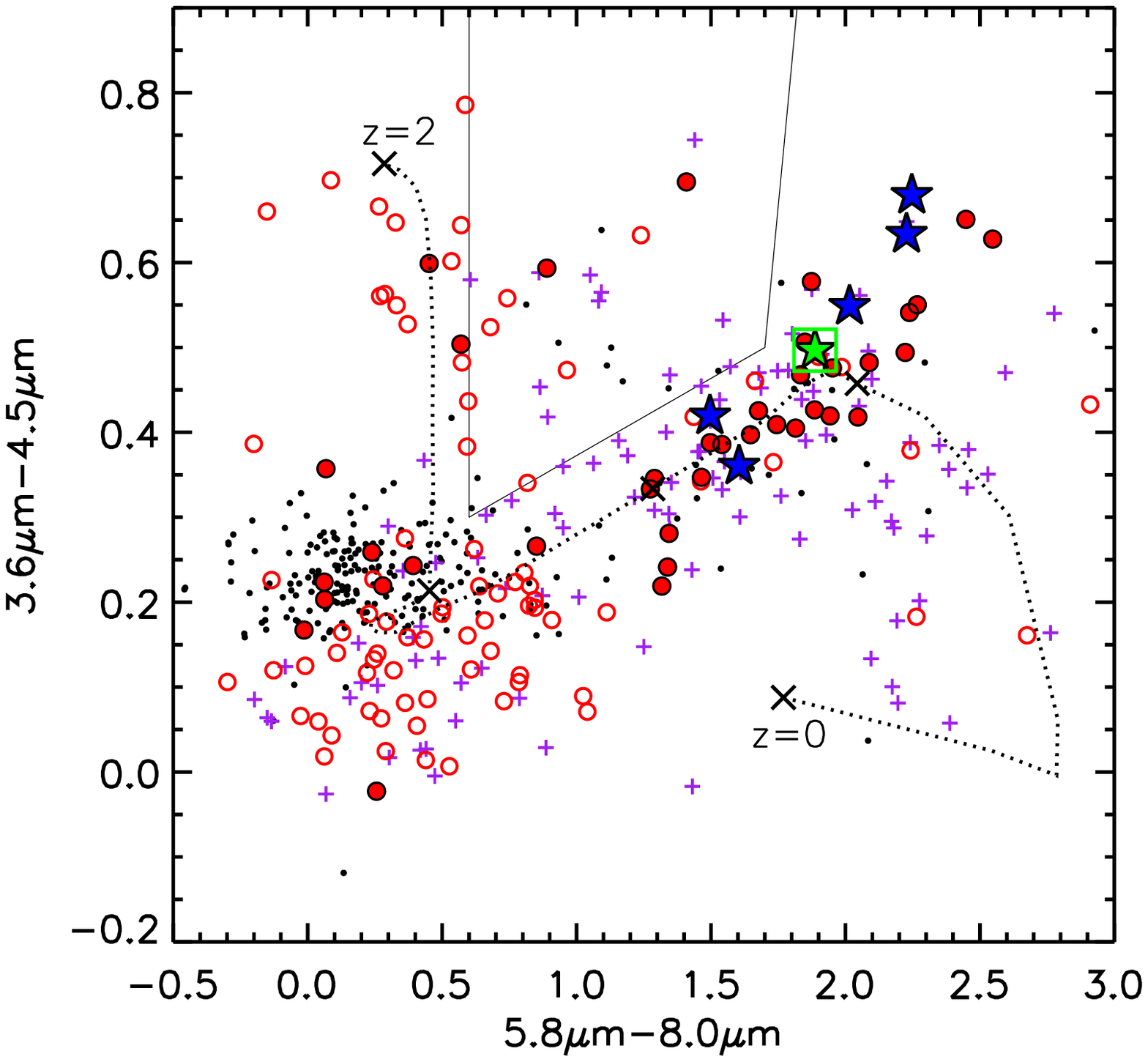}{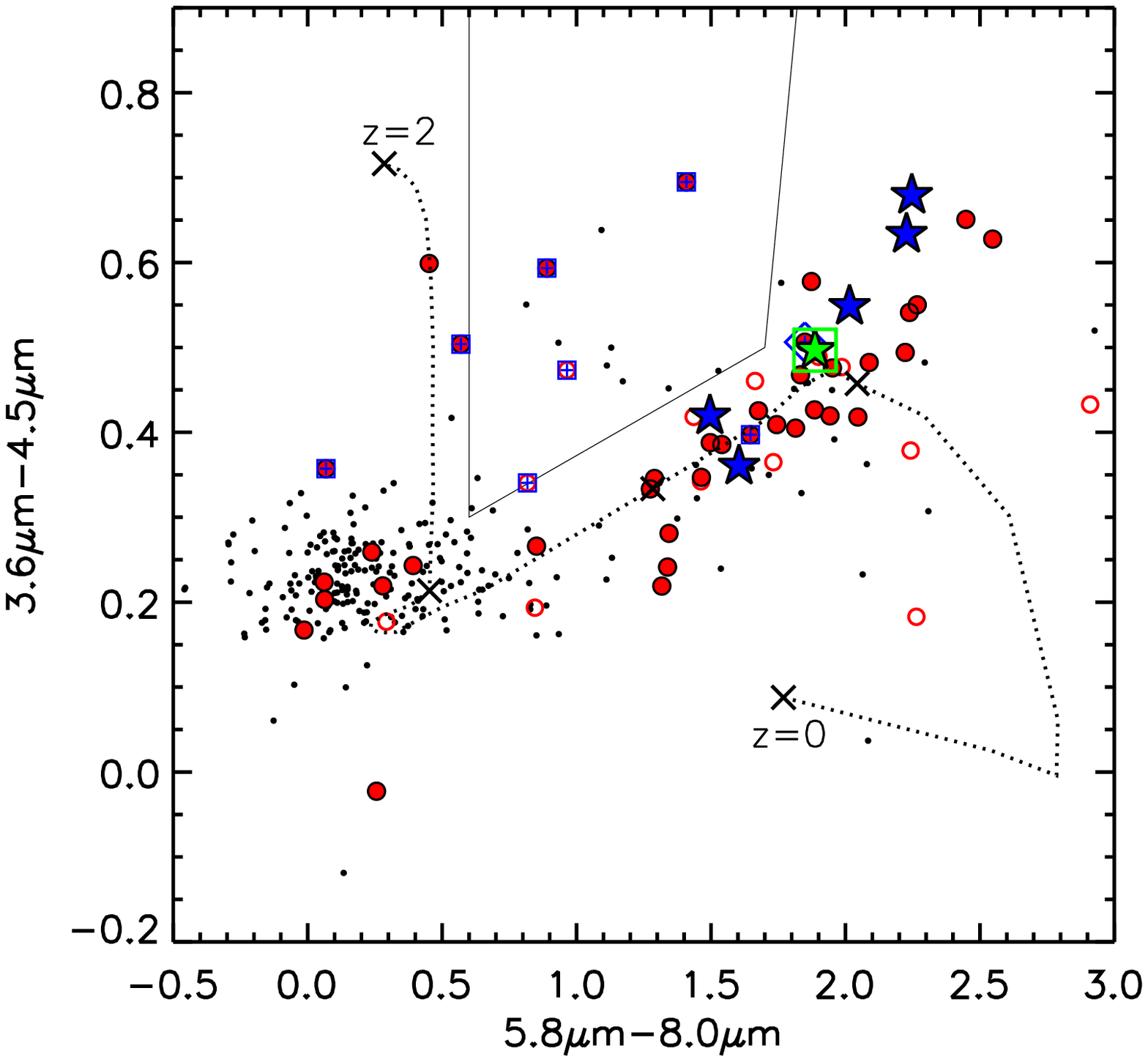}
\caption{IRAC color-color diagram, showing the AGN wedge \citep{lacy2004,stern2005}.  Symbols are the same as for Figure~\ref{fig:select}.  MIPS sources identified as IRAC or X-ray AGN are shown with a blue square and plus sign overplotted on a solid or open red circle.  The Seyfert galaxy identified in Figure~\ref{fig:BPT} is shown with a blue diamond behind the ULIRG (green star symbol). The dotted line shows the M82 track starting from $z=0$ then progressing to $z=0.3, 0.5, 1,$ and 2, with each redshift marked by a cross.  It is clear that many of the high redshift background sources are removed from our color selection based on Figure~\ref{fig:select}.}
\label{fig:irac_colors}
\end{figure*}

We then use IRAC colors to identify AGN via the ``AGN wedge'' \citep{lacy2004,stern2005}, as shown in Figure~\ref{fig:irac_colors}.  We exclude all galaxies within the AGN wedge from further analyses because their mid-infrared luminosity may be dominated by AGN activity rather than by star formation.  We also exclude X-ray AGN using a catalog of 145 X-ray point sources extracted from Chandra data that cover the central $\sim20\arcmin\times 20\arcmin$, overlapping the entire MIPS FOV.  The catalog contains X-ray sources down to a flux of  $2.5\times10^{-16}$ ergs cm$^{-2}$ s$^{-1}$ in the 0.5-2 keV band, which is a luminosity of $L_{x}=7\times10^{40}$ ergs s$^{-1}$ at the redshift of the Bullet Cluster.  Among the MIPS confirmed cluster members and candidates, we exclude all X-ray point sources with an X-ray luminosity of $L_{x}\geq10^{41}$ ergs s$^{-1}$.

\begin{figure*}
\epsscale{1} 
\plottwo{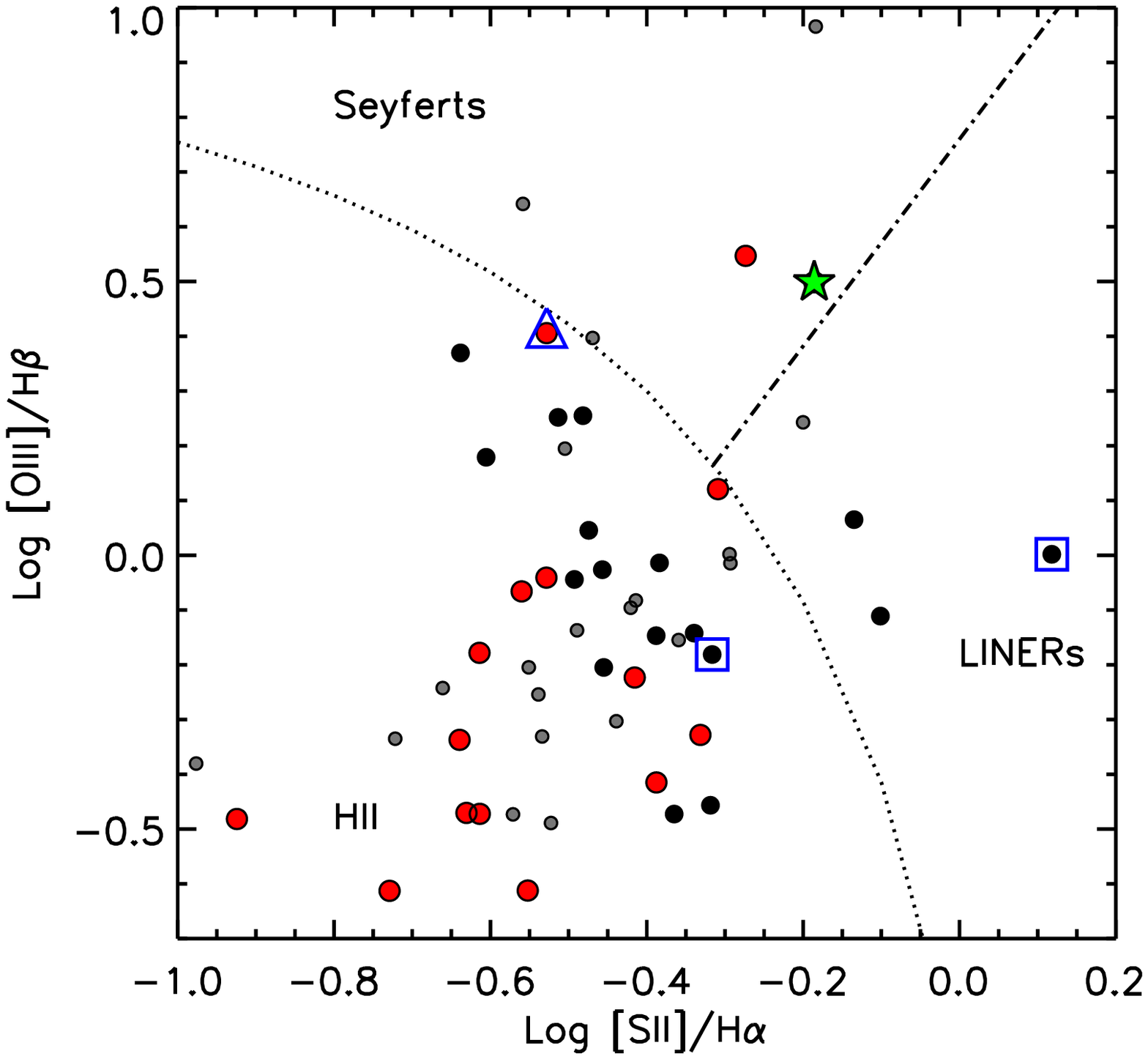}{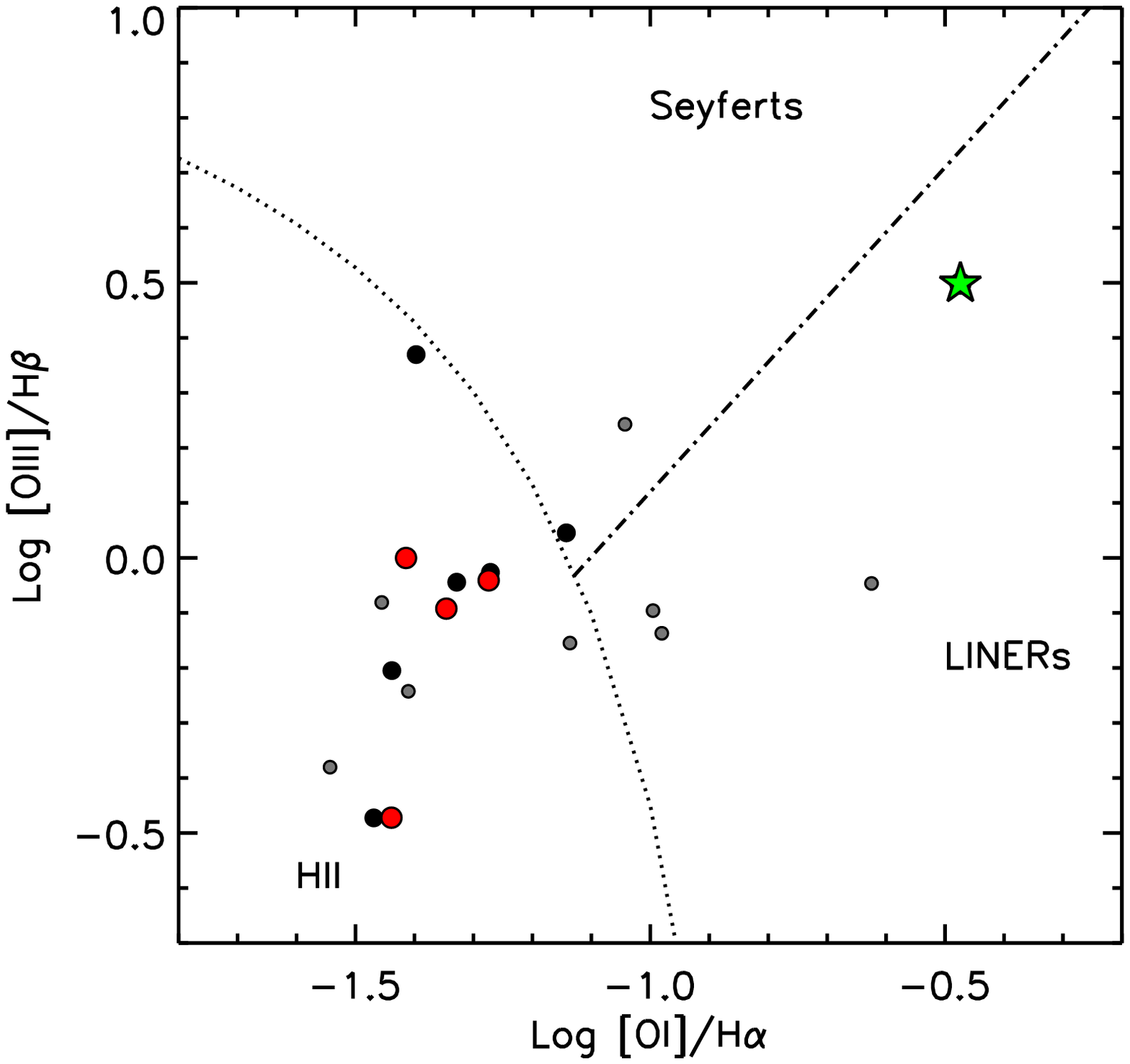} 
\caption{BPT diagrams with dotted and dashed lines separating purely star-forming galaxies from Seyferts and LINERs \citep{kewley2006}. Small gray circles are objects that are beyond the IRAC FOV, and therefore it is unknown whether they would fall in the IRAC AGN wedge.  Large circles represent objects within the IRAC FOV, with red (gray) and black circles indicating galaxies with and without 24\micron\ emission, respectively.  Objects outlined with a large square and a large triangle have been identified as AGN via IRAC colors and X-ray emission, respectively.  The star symbol represents the ULIRG, which falls close to the Seyfert/LINER boundary in the left panel, and is classified as a LINER in the right panel.} 
\label{fig:BPT} 
\end{figure*}

In addition to IRAC and X-ray selection of AGN, we also utilize our optical spectroscopy to construct Baldwin-Phillips-Terlevich diagrams \cite[BPT;][]{baldwin1981} as illustrated in Figure~\ref{fig:BPT}.  The dotted and dashed lines indicate the boundaries from \citet{kewley2006} that separate purely star forming galaxies from Seyferts and LINERs.  Note that many of the MIPS sources are missing in Figure~\ref{fig:BPT} because one or more of the four required emission lines could not be measured, often due to a prominent sky emission line that appears at the same wavelength as H$\beta$ at the Bullet Cluster redshift.  

Figure~\ref{fig:BPT} reveals only two MIPS sources classified as a Seyfert or LINER that have not been identified as AGN using either IRAC colors or X-ray emission.  One of these is a ULIRG (star symbol), which appears close to the Seyfert/LINER boundary when using the [SII]/H$\alpha$ ratio, and is classified as a LINER using the [OI]/H$\alpha$ ratio.  

The two confirmed MIPS sources in the IRAC AGN wedge (Figure~\ref{fig:irac_colors}) are not shown on the BPT diagram because they lack the necessary emission lines.  However, of the 10 confirmed IRAC AGN wedge sources that are not MIPS members (or beyond the MIPS FOV), two are plotted in Figure~\ref{fig:BPT}, with one classified as a LINER, and the other as an HII dominated galaxy.

Of the three X-ray point sources in Figure~\ref{fig:irac_colors}, one is illustrated in Figure~\ref{fig:BPT} as a large square and appears close to the boundary of HII dominated galaxies and Seyferts.

In total, there are 8 AGN identified among the confirmed MIPS sample, using the three methods -- IRAC colors, X-ray emission, and optical emission line ratios.  Of these eight, two are identified solely from the BPT diagnostic, one classified as a LINER and the other a Seyfert.  The LINER is included in our sample, assuming that 65\% of its IR flux is powered by star formation (see \S\ref{sec:ULIRG}).  The Seyfert, which is excluded from our sample, would contribute a neglible fraction to the global SFR.

In addition to isolating the AGN population, Figure~\ref{fig:irac_colors} also illustrates that our selection of cluster candidates based on R-[4.5] and B-R colors is an effective way to cull interlopers.  The left and right panels of Figure~\ref{fig:irac_colors} show the IRAC color distribution before and after we apply the  R-[4.5] and B-R color selection, respectively.  We show the model colors of M82 (a local starburst galaxy) at z=0,0.3,0.5, and 1 \citep{devriendt1999,stern2005} to illustrate that our color selection is effective in removing galaxies whose IRAC colors are consistent with those of a high redshift starburst (open red circles near $z\sim1$ to $2$).  

The right panel of Figure~\ref{fig:irac_colors} highlights the AGN sources among the MIPS sample, including four galaxies in the IRAC AGN wedge, three X-ray point sources outside the AGN wedge, and one Seyfert identified from the BPT diagram, which is semi-hidden behind the ULIRG symbol.  Among the non-AGN MIPS cluster members, there are two outliers, both of which do not fit in the diagonal star forming sequence (roughly outlined by the LIRGs and ULIRG shown as star symbols) nor the locus of passive early type galaxies near [3.6]-[4.5]$\sim$0.2 and [5.8]-[8.0]$\sim$0.2.  One outlier has a color of [3.6]-[4.5]$<0$, due to a blending of sources in the 3.6\micron\ and 4.5\micron\ bands.  It is unclear why the second outlier has IRAC colors consistent with those of a high redshift starburst. Perhaps an AGN component undetected in X-ray is contaminating its IRAC colors.  We include both sources in the following analyses since they are cluster members with 24\micron\ emission and have no sign of a dominant AGN component.  Both galaxies have a SFR$<3.6$ \Msolaryr\ and contribute negligibly to any results presented in this paper.

\subsection{Spectroscopic Completeness}
We match our photometric catalogs to our spectroscopic one, which includes 857 redshifts within the full IMACS FOV ($25^{'}\times 25^{'}$), centered on the Bullet Cluster (Figure~\ref{fig:veldisp}).   Of these, 362 galaxies are confirmed cluster members and 495 are interlopers.  Combined with the \citet{barrena2002} catalog, we have a total of 406 confirmed cluster members.

In estimating the spectroscopic completeness of MIPS galaxies, we consider only the 139 MIPS sources that lie within the diagonal color boundaries in Figure~\ref{fig:select}.  Out of those, nine lie within the AGN wedge -- two confirmed members, five interlopers, and two cluster candidates.  The spectroscopic completeness within the AGN wedge for MIPS sources is $\sim$80\%.  

Out of the 130 MIPS galaxies that lie outside the AGN wedge, we have redshifts for 118 galaxies, with 39 confirmed cluster members, 78 interlopers, and 12 cluster candidates with no spectroscopy.  We are 90\% spectroscopically complete for MIPS sources outside the AGN wedge, with $\sim$35\% being cluster members.

There are a total of 43 confirmed members with MIPS emission, including two cluster members whose R-[4.5] colors are slightly beyond the color selection of Figure~\ref{fig:select}, and five members whose redshifts are from \citet{barrena2002}.  Among these 43, we exclude two in the IRAC AGN wedge, three that are X-ray point sources, and one that is a Seyfert.  Of the additional 14 MIPS cluster candidates chosen based on IRAC and optical colors, two are in the AGN wedge and hence excluded.  Our final non-AGN MIPS sample consists of 37 confirmed members and 12 candidate members, of which we expect $\sim$4 to be members based on our success rate for targeting MIPS cluster members.

\subsection{Total Infrared Luminosity}

To determine the total infrared luminosity $L_{IR}$ ($\lambda=8-1000$ \micron) and star formation rate (SFR) from the observed 24\micron\ flux $f(24)$, we use the prescription in \citet{rieke2009}, who construct model spectral energy diagrams (SEDs) for a sample of local LIRGs and ULIRGs and derive relations between 24\micron\ flux, star formation rate, and total infrared luminosity.  The scatter associated with the $L_{24}-L_{IR}$ relation is 0.13 dex.  

\citet{rawle2010} have found that for a sample of 23 Bullet Cluster galaxies, $\sim$30\% are found to have an excess of 100\micron\ flux relative to the \citet{rieke2009} and \citet{dale2002} templates.  Although the deviation between {\it Herschel} derived and 24\micron\ derived total infrared luminosities can be up to a factor of 4 for some sources, these are typically low-luminosity sources fainter than our S/N criterion for the 24 micron sample.  Among the galaxies used in this paper (with available {\it Herschel} data), the ratio of SFR$_{FIR}$ to SFR$_{24}$ is close to 1, with a scatter of 0.6, indicating that the 24\micron\ flux yields accurate total IR luminosities and star formation rates for our sample.

\begin{figure*}
\epsscale{1.}
\plottwo{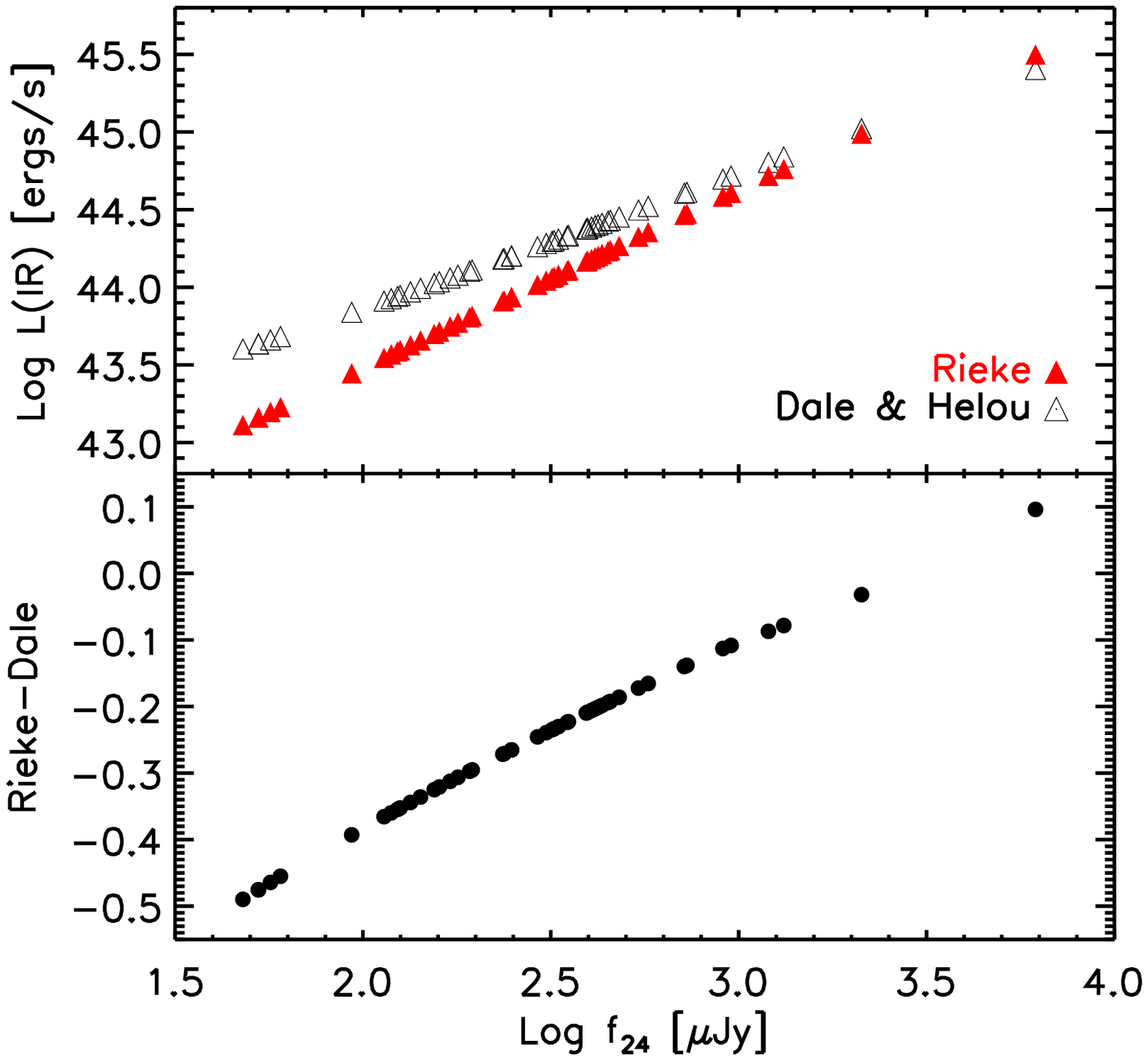}{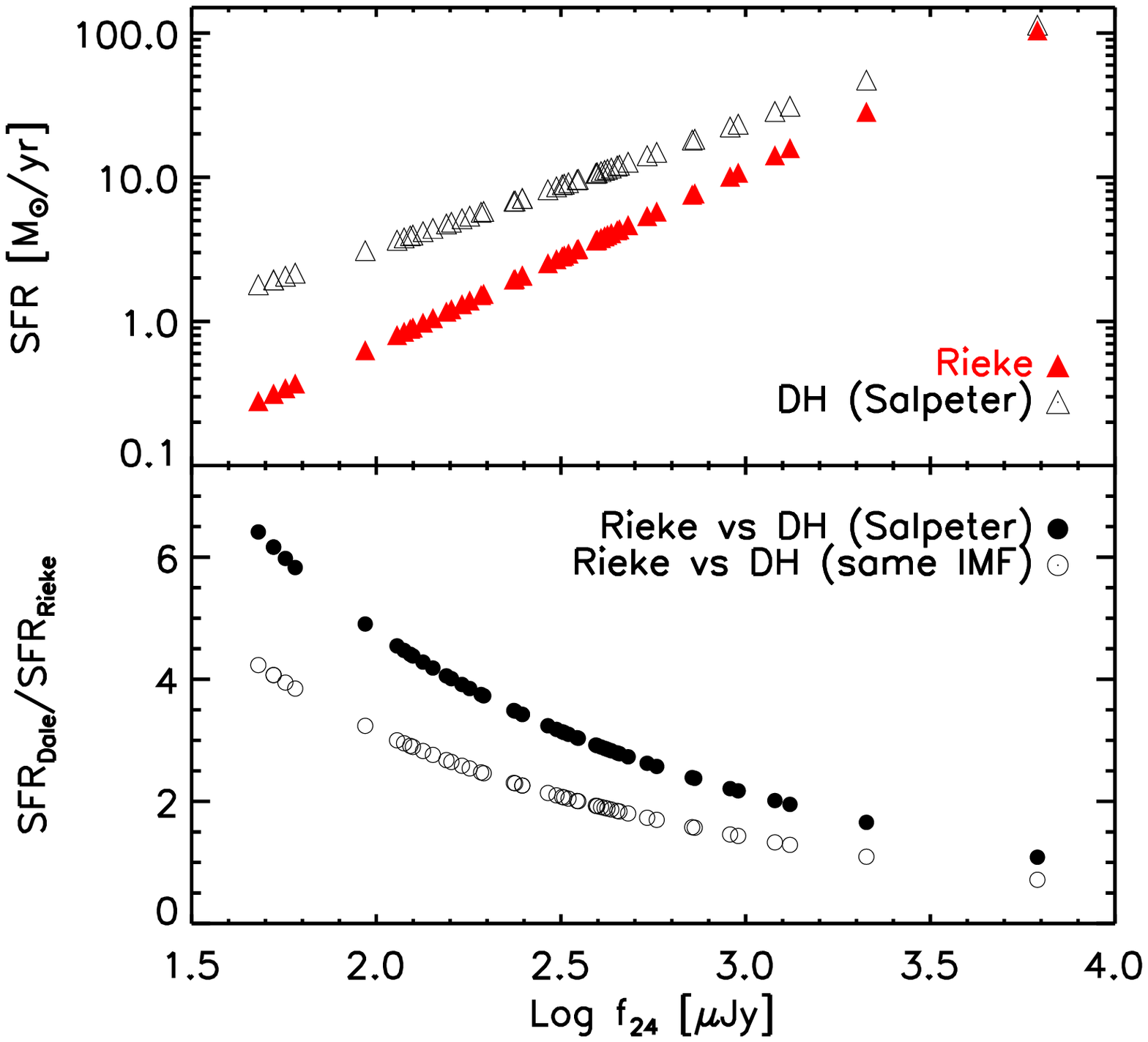} 
\caption{The left and right panels show a comparison of the total infrared luminosity and star formation rate derived from the \citet{rieke2009} relation (red triangles) compared to those from the \citet{dale2002} templates with the \citet{kennicutt1998} star formation relation (black circles), as a function of 24\micron\ flux for the 49 non-AGN MIPs confirmed members and cluster candidates in our sample.  The bottom portion of both panels show the systematic offsets between the two methods.} 
\label{fig:dale_rieke}
\end{figure*}

Although we have a decent understanding of the systematic uncertainty in the $L_{IR}$ derived from the \citet{rieke2009} templates, much of the previous work in the literature uses the galaxy templates of \citet{dale2002} to derive $L_{IR}$ from observed 24\micron\ flux.  In the left panel of Figure~\ref{fig:dale_rieke}, we show the relation between $L_{IR}$ and observed 24\micron\ flux derived from the \citet{dale2002} templates at z=0.3 for our 49 non-AGN MIPS members and cluster candidates.  We compare the \citet{dale2002} $L_{IR}$ values to those derived using the \citet{rieke2009} calibration.  The lower portion of the left panel in Figure~\ref{fig:dale_rieke} shows the difference between the two sets of $L_{IR}$ as a function of observed 24\micron\ flux.

Among the 49 MIPS sources, the \citet{rieke2009} and \citet{dale2002} templates yield a difference in total infrared luminosity of approximately -0.5 dex, -0.25 dex, and 0.1 dex for the low, median, and high end of 24\micron\ fluxes in our sample, respectively.  The brightest 24\micron\ source shown in Figure ~\ref{fig:dale_rieke} is 65\% of the ULIRG flux.

\subsection{Star Formation Rate}
\label{sec:sfr_systematics}

We calculate the star formation rate from the observed 24\micron\ flux by using equation~14 in \citet{rieke2009}.  The uncertainty in the SFR derived from this relation is $\sim$0.2 dex.  This error is dominated by the scatter in the $L_{24} - L_{IR}$ relation, with an additional error due to the underlying assumption that most of the young stellar light is absorbed and re-radiated in the infrared.  Although an average correction has been applied to account for the loss of young stellar light directly to the ultraviolet (and therefore never seen in the infrared), there is a variation from galaxy to galaxy that adds an additional uncertainty to the star formation rate calibration.  The 0.2 dex estimate does not include uncertainties in the assumed initial mass function or the theoretical relation between SFR and $L_{IR}$.  We assign an error of 0.2 dex to all star formation rates calculated in this paper.  

In addition to the uncertainty in the star formation rate, we need to understand any systematic offsets between the SFRs calculated from the \citet{rieke2009} relations, versus the SFRs quoted in much of the previous literature using the \citet{dale2002} galaxy templates and the original \citet{kennicutt1998} SFR-$L_{IR}$ relation.  Star formation rates obtained from the SFR-$L_{IR}$ relation presented in \citet{rieke2009} are systematically lower by $\sim$0.2 dex relative to the standard \citet{kennicutt1998} relation due to a difference in the initial mass function (IMF).  An unbroken Salpeter IMF of slope -1.35 from 0.1 to 100 \Msun\ is used in the original \citet{kennicutt1998} derivation, whereas more recent studies suggest that a Salpeter-like IMF with a shallower slope for the low mass end is more applicable for extragalactic star-forming regions \citep{rieke1993,alonso2001}.  This broken Salpeter IMF has a total mass that is $\sim$0.66 times the mass of a single power-law Salpeter IMF, and yields a similar proportion of stars to the \citet{kroupa2002} IMF and the \citet{chabrier2003} IMF.

The right panel of Figure~\ref{fig:dale_rieke} shows the comparison of SFRs derived from the \citet{rieke2009} calibration compared to those from the \citet{dale2002} templates and \citet{kennicutt1998} relation, as a function of the observed 24\micron\ flux of the Bullet Cluster MIPS sources.  The SFRs calculated from the two methods are systematically offset from each other more significantly than in the case of total infrared luminosities.  The bottom portion of the right panel of Figure~\ref{fig:dale_rieke} shows that the \citet{rieke2009} calibration yields systematically lower star formation rates by a factor of $\sim$2.6 to 1.2 for the range of 24\micron\ fluxes in the Bullet Cluster sample, as noted by the dotted vertical lines.  The difference in SFRs derived from the two methods is in part due to the systematically lower infrared luminosities derived from the the \citet{rieke2009} templates, as well as the assumption of a different IMF from the \citet{kennicutt1998} relation which produces fewer low mass stars.

\section{Results \& Discussion}

\subsection{LIRGs and ULIRG}
\label{sec:ULIRG}
Among the 37 non-AGN MIPS confirmed cluster members, five are LIRGs ($L_{IR}>10^{11}$\Lsun) and one is a ULIRG ($L_{IR}>10^{12}$\Lsun).   Although the ULIRG does not appear within the AGN wedge nor as an X-ray point source, its optical spectrum exhibits broad emission lines and emission line ratios on the BPT diagrams \cite[BPT;][]{baldwin1981} that classify this galaxy as a borderline Seyfert/LINER when using the [SII]/H$\alpha$ ratio and a LINER with the [OI]/H$\alpha$ ratio (Figure~\ref{fig:BPT}).  

Mid-infrared and far-infrared emission in ULIRGs arises from dust that is heated by a young stellar population and/or a central AGN.  In the case of LINERs, some studies have found that the infrared emission is dominated by star formation rather than nuclear activity \cite[e.g.][]{satyapal2005,veilleux2009}.  Using six different methods to determine AGN contribution, \citet{veilleux2009} found that the AGN contribution to the bolometric luminosity in LINERs is $\sim$15\% to 35\%.  In addition, \citet{sturm2006} found that IR-bright LINERs tend to have mid-IR SEDs consistent with starbursts, whereas SEDs of IR-faint LINERs appear more AGN dominated.  The LINER in our sample has a total IR luminosity of $\sim10^{12}$ \Lsun, making it more likely to be starburst dominated.  While we do not know the precise AGN contribution in our ULIRG, we adopt a value of 35\%, in a cautious effort not to overestimate the true star formation rate.   In the proceeding analyses and discussion, we include this ULIRG in our sample, but always with a star formation rate that assumes that only 65\% of the infrared flux is powered by star formation.  The sky coordinates of the five LIRGS and one ULIRG are listed in Table~\ref{table:LIRGs}, along with their total infrared luminosities and star formation rates.

\begin{figure}
\epsscale{1}
\plotone{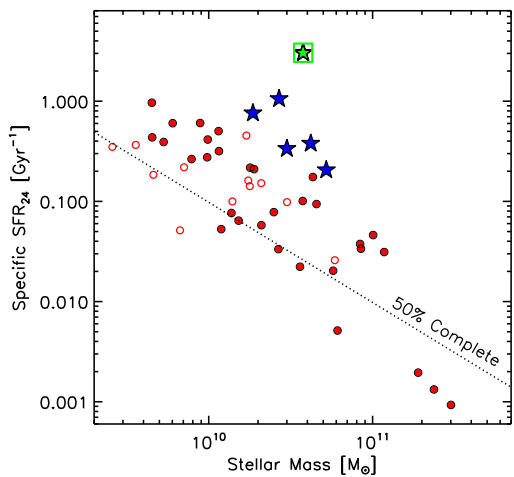}
\caption{The star formation rate per stellar mass (specific star formation rate) as a function of stellar mass, with the 50\% SFR completeness limit indicated (dotted line).  The five LIRGs and one ULIRG are highlighted as star symbols, with a square border around the ULIRG.}
\label{fig:SSFR}
\end{figure}

Figure~\ref{fig:SSFR} shows the specific star formation rate (SSFR) and stellar mass for the LIRG and ULIRG population, and the remaining non-AGN MIPS sample of 37 confirmed members and 12 cluster candidates.  There is a clear trend of decreasing specific SFR with increasing stellar mass.  Although there is an inherent selection bias against low mass, low specific SFR galaxies, there is evidence that the general correlation seen in Figure~\ref{fig:SSFR} is real.  The upper envelope of galaxies with high SFRs (parallel to and above the 50\% completeness line in Figure~\ref{fig:SSFR}) should represent a nearly complete sample of bright 24\micron\ sources.  Among this nearly complete sample of high SFR galaxies, the trend of decreasing specific SFR as a function of stellar mass is still apparent.  In addition, the correlation seen in Figure~\ref{fig:SSFR} has been observed in several other studies, in both field and cluster galaxies \citep{feulner2005,perezgonzalez2005,noeske2007,vulcani2010}.  We note that Figure~\ref{fig:SSFR} is intended to illustrate a general trend between SSFR and stellar mass (not a precise fit or relation), as well as the range of SSFR and stellar mass values in our MIPS sample.

The five LIRGs and ULIRG, which are highlighted (star symbols) in Figure~\ref{fig:SSFR}, represent an outlier population among the MIPS sources.  They show excess levels of star formation given their stellar mass, with the ULIRG being the most extreme outlier from the nominal SSFR versus stellar mass relation.

\begin{figure}
\epsscale{1}
\plotone{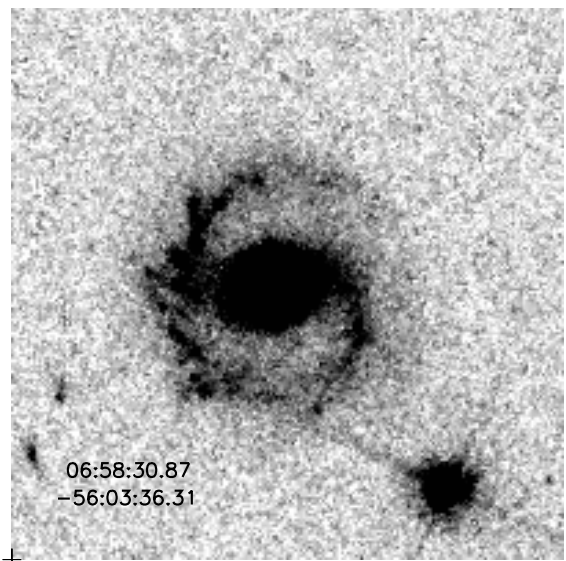} 
\caption{The ULIRG imaged with HST/ACS in the F606W filter.  It is a barred spiral with no signs of recent major interactions.}
\label{fig:HST}
\end{figure}

All galaxies listed in Table~\ref{table:LIRGs} lie beyond R$\sim$1 Mpc from the cluster center, with the two most luminous galaxies located at R$\sim$1.7 Mpc.  The intermediate to cluster outskirts region is an environment similar to that of galaxy groups, where galaxy-galaxy interactions are more likely to occur than in high density cluster cores or in the low density field.  However, the visual morphologies of the LIRGs do not show any obvious signs of recent merger activity, based on BVR imaging from the Magellan 6.5m Baade telescope.  At least two out of the five LIRGs are clearly spiral galaxies, while morphologies of the other three are not entirely clear, though they do show some faint spiral structure.  Figure~\ref{fig:HST} shows an image of the ULIRG taken with the {\it Advanced Camera for Surveys} on the {\it Hubble Space Telescope} in the F606W filter (P.I. Holland Ford; Proposal ID 10996).  It is a barred spiral galaxy, with no obvious signs of recent interaction or mergers, though there is some assymetry in the spiral arms.

\begin{figure}
\epsscale{1}
\plotone{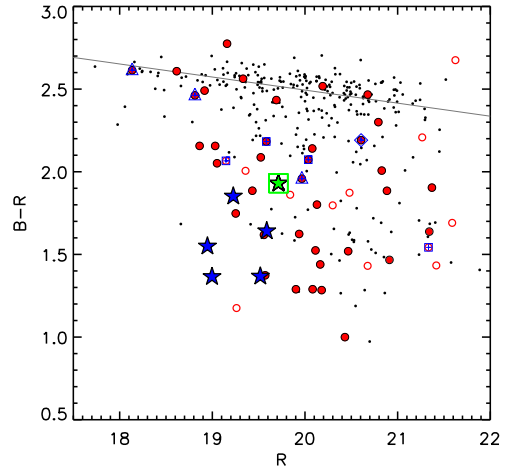}
\caption{The optical color-magnitude diagram with spectroscopically confirmed cluster members (dot), spectroscopically confirmed members with MIPS emission (solid circle) and cluster candidates with MIPS emission (open circle).  MIPS sources identified as AGN with IRAC, X-ray, or optical data are indicated with a blue square, triangle, and diamond with a plus sign, respectively. Five spectroscopically confirmed LIRGs are shown as blue stars, and one ULIRG/LINER as a green star with a square border.  Grey solid line shows the color-magnitude relation with slope and normalization adopted from \citet{lopez-cruz2004}.}
\label{fig:CMD}
\end{figure}

All six LIRGs/ULIRG have optical colors consistent with those of star-forming, blue cloud galaxies, as shown in the color magnitude diagram in Figure~\ref{fig:CMD}.  The IRAC color-color diagram (Figure~\ref{fig:irac_colors}) also shows that the LIRGS and ULIRG  have colors consistent with star-forming galaxies, with the ULIRG having a similar color as the starburst galaxy M82 evolved to $z=0.3$.

Table~\ref{table:LIRGs} also lists the six cluster members that are classified as AGN from IRAC, X-ray, and optical spectroscopic data.  One out of the six galaxies is an X-ray point source and LIRG, with $L_{IR}\sim 10^{11}$ \Lsun, lying close to the IRAC AGN wedge.  Optical colors of the six confirmed and two candidate member AGN are shown in Figure~\ref{fig:CMD}. 

\subsection{Global Star Formation Rate}
\label{sec:globalSFR}

The integrated SFR of the 37 MIPS confirmed cluster members within $R<1.7$ Mpc is 267 \Msolaryr, with a 0.2 dex uncertainty (see \S\ref{sec:sfr_systematics}).  The 12 MIPS candidate members contribute an additional 27 \Msolaryr.  Normalizing by the cluster mass $M(<1.7Mpc)=9.5\times 10^{14}$ \Msun, derived from the caustic analysis of the cluster infall region (Gonzalez et al. in prep), we obtain 28 \Msolaryr\ per $10^{14}$ \Msun\ for the confirmed members sample, with the ULIRG contributing 40\% and the five LIRGs an additional 30\% (see Table~\ref{table:LIRGs}).  

\begin{figure}
\epsscale{1}
\plotone{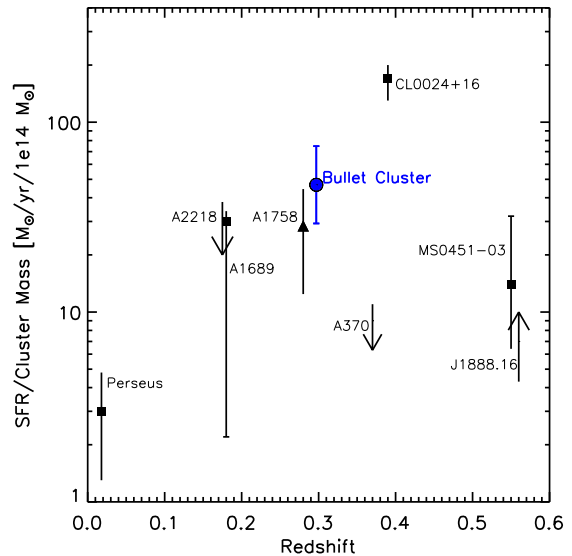}
\caption{The mass normalized integrated SFR as a function of redshift, with the Bullet Cluster shown as a blue solid circle.   SFRs calculated by \citet{haines2009} for A1758 is illustrated with a triangle. All others are calculated in \citet{geach2006} and references therein.}
\label{fig:SFR}
\end{figure}

The Bullet Cluster integrated star formation rate normalized by cluster mass is presented in Figure ~\ref{fig:SFR}, along with known values for eight other clusters taken from \citet{geach2006} and references therein, and \citet{haines2009}, who use 24\micron\ MIPS or 15\micron\ ISOCAM data to calculate obscured SFRs.  In order to facilitate a fair comparison of SFR in the Bullet Cluster and the other clusters shown in Figure~\ref{fig:SFR}, we impose an infrared luminosity lower limit on our MIPS sample.  In addition, we scale the star formation rate to account for a systematic offset in the \citet{rieke2009} relations versus the \citet{dale2002} templates with the original \citet{kennicutt1998} relation (Figure~\ref{fig:dale_rieke}).  

\citet{geach2006} apply a lower limit of $L_{IR}=6\times 10^{10}$ \Lsun\ ($SFR\sim 10$ \Msolaryr), for all their clusters with the exception of MS0451-03.  This lower limit was calculated for a 200 $\mu$Jy source at $z=0.39$ using the \citet{dale2002} templates.  Using the \citet{rieke2009} calibration, the same source at $z=0.39$ yields a total IR luminosity of $L_{IR}=3.8\times 10^{10}$ \Lsun\ ($SFR=3.6$ \Msolaryr).  After applying a lower limit of $L_{IR}=3.8\times 10^{10}$ \Lsun to our full sample of 37 confirmed members and 12 cluster candidates, 20 galaxies remain -- 19 members and one candidate.  We then scale the SFR of each galaxy  using the relation presented in the bottom right panel of Figure~\ref{fig:dale_rieke}, in order to convert SFRs calculated from the \citet{rieke2009} relation to the \citet{dale2002} system used by \citet{geach2006}.  After making these adjustments,  we obtain an integrated SFR of 445 \Msolaryr\ for the 20 galaxies, and a cluster mass normalized SFR of 47 \Msolaryr\ per $10^{14}$ \Msun, as shown in Figure~\ref{fig:SFR}.

All star formation rates illustrated in Figure~\ref{fig:SFR} are derived from mid-infrared data, primarily with the MIPS 24\micron\ band and in some cases supplemented with data from the Infrared Satellite Observatory (ISO) 15\micron\ band.  Integrated SFRs calculated by \citet{geach2006} include mid-IR sources out to a cluster radius of $R<2$ Mpc, while the SFR of A1758 is obtained from mid-IR data out to $R<3$ Mpc (\citet{haines2009}).  This is comparable to the $R<1.7$ Mpc area that we survey. 

The mass normalized SFR of the Bullet Cluster presented in this paper is a more precise estimate of the SFR than what was possible for many of the clusters shown in Figure~\ref{fig:SFR}, which mostly lacked mid-infrared data beyond the cluster core.  Extrapolation of mid-infrared sources and field subtraction were applied in such cases, to place upper and lower limits on the global specific SFR, illustrated as horizontal bars and arrows on Figure~\ref{fig:SFR}.  In addition, the spectroscopy was sparse for several of the clusters presented in Figure~\ref{fig:SFR}, whereas the Bullet Cluster SFR is based on a 90\% spectroscopically complete MIPS sample.

While the integrated SFR of the Bullet Cluster excludes contamination from both mid-IR and X-ray AGN, many of the clusters considered by \citet{geach2006} and \citet{haines2009} lack the necessary data for a similarly complete AGN removal. In A1758, \citet{haines2009} identify two X-ray point sources that are removed as AGN, though no consideration is made for mid-infrared AGN without X-ray emission.  \citet{geach2006} remove AGN from CL0024+16 and MS0451-03 by excluding mid-IR sources that have optical and K-band colors similar to the expected colors of passive early type galaxies (E/S0) from \citet{king1985} models.  Although this may help eliminate AGN with early type hosts, there are two main issues with using this color criterion to select out AGN.

The first problem is that identifying AGN solely with this method can be significantly incomplete. As our own data show in Figure~\ref{fig:irac_colors}, none of the AGN (six confirmed members, two cluster candidates) have IRAC colors consistent with the locus of early type galaxies near [3.6]-[4.5]$\sim$0.2 and [5.8]-[8.0]$\sim$0.2.  In Figure~\ref{fig:CMD}, the optical color-magnitude diagram shows that only two out of the eight AGN are on the red sequence.  

The second problem is that attempting to select out early-type AGN hosts based on their optical and K-band colors can inadvertently remove many dusty star-forming galaxies, which may actually be spiral galaxies with optically red colors.  Figure~\ref{fig:CMD} shows that eight out of the 37 non-AGN MIPS cluster members lie on the red sequence.  Although 20\% of the non-AGN MIPS cluster members are optically red, these galaxies have a total SFR of $\sim$14 \Msolaryr, contributing only $\sim$5\% to the integrated SFR.  While the global SFR in the Bullet Cluster is dominated by LIRGs/ULIRG, optically red mid-IR galaxies may have more of an impact on the global SFR in clusters that lack a LIRG/ULIRG population.  In general, it is important not to exclude galaxies as AGN based purely on their optical colors, as we have seen that up to 20\% of star-forming galaxies can lie on the red sequence.

Had we used a similar criterion as \citet{geach2006} to exclude AGN based on IRAC and optical colors of passive galaxies, we would not have detected any of the six confirmed or two candidate AGN, whose total infrared luminosity converts to a star formation rate of 42 \Msolaryr\ for the confirmed members, and 8 \Msolaryr\ for the two cluster candidates, using the \citet{rieke2009} calibration.  Adjusting for the different SFR calibration used in \citet{geach2006}, the global specific SFR of the Bullet Cluster would increase from 47 to 57 \Msolaryr\ per $10^{14}$ \Msun\ in Figure~\ref{fig:SFR}. 

\subsection{Specific SFR in the Bullet Cluster}

The mass normalized integrated SFR of the Bullet Cluster is among the highest in the sample of local to intermediate redshift clusters shown in Figure~\ref{fig:SFR}, second only to CL0024+16.  The specific SFR of the Bullet Cluster is also comparable to that of A1758, a cluster merger at nearly the same redshift.  Interestingly, the three most active clusters (CL0024+16, A1758, and the Bullet Cluster) in Figure~\ref{fig:SFR} are all cluster mergers.  However, all three are in different merger states, with different masses and dynamical histories.  Although Figure~\ref{fig:SFR} is somewhat suggestive that cluster mergers have elevated global specific SFRs compared to non-merging clusters, a larger sample of clusters is necessary to confirm a trend.  

Previous studies from \citet{kodama2004,geach2006,bai2007,koyama2010} have suggested that there is a redshift dependence on the global star formation rate of clusters, anywhere from $(1+z)^{4}$ to $(1+z)^{7}$, based on plots similar to Figure~\ref{fig:SFR}.  Although Figure~\ref{fig:SFR} shows some suggestion of evolution in the specific SFRs, there is clearly much scatter between individual clusters.  We emphasize that the integrated SFR is highly sensitive to small number statistics.  In the case of the Bullet Cluster, one galaxy is contributing $\sim$40\% of the entire integrated SFR.  
 
Another way to compare the star formation activity in different clusters is to look at the mass normalized integrated SFR as a function of cluster mass.  As found in \citet{bai2007,koyama2010}, there is a dependence of specific SFR on total cluster mass, although it does not display a much stronger correlation than with redshift. 

Of all the clusters in Figure~\ref{fig:SFR}, CL0024+16 and the Bullet Cluster have the highest number of spectroscopically confirmed 24\micron\ MIPS sources, with 45 and 44 cluster members respectively, prior to excluding AGN.  Although CL0024+16 has a specific SFR that is a factor of $\sim$4 higher than the Bullet Cluster, this could in part be driven by its small cluster mass \cite[$M(<2 Mpc)=6.1\times 10^{14}$;][]{kneib2003}, as several authors have noted a correlation between global specific SFR and cluster mass \cite[e.g.][]{finn2005,bai2007,koyama2010}.  

Among the three clusters with highest specific SFR (A1758, Bullet Cluster, and CL0024+16), we note that the correction for AGN contamination in CL0024+16 and A1758 is less complete than for the Bullet Cluster, as is the spectroscopic completeness of A1758.  In addition, the SFR for A1758 is integrated from candidate members out to $R<3$ Mpc, compared to $R<1.7$ Mpc for the Bullet Cluster.  It has been shown in several studies that infrared and submm luminous galaxies with high levels of star formation rate are preferentially distributed in cluster outskirts \citep[e.g.][]{ma2010,braglia2010}. Therefore a comparison of integrated SFR (not normalized by mass), within $R<3$ Mpc versus $R<1.7$ Mpc is not necessarily a fair view on the relative star formation activity between two clusters.  

The integrated SFR of the Bullet Cluster is strongly driven by a small number of infrared luminous galaxies.  Among our sample of 37 MIPS confirmed members excluding AGN, one ULIRG contributes 40\% of the total SFR, while five LIRGs contribute an additional 30\%.  We will see in the subsequent sections that these few galaxies not only dominate the global SFR, but have a significant impact on the infrared luminosity function as well.

\subsection{Infrared Luminosity Function}
\label{sec:IRLF}

To further investigate the origin of the Bullet Cluster elevated global star formation rate relative to other clusters, we construct an infrared luminosity function of the star-forming population and compare it to the IR LF of non-merging clusters with similar redshift and mass.  

\begin{figure*}
\epsscale{1}
\plottwo{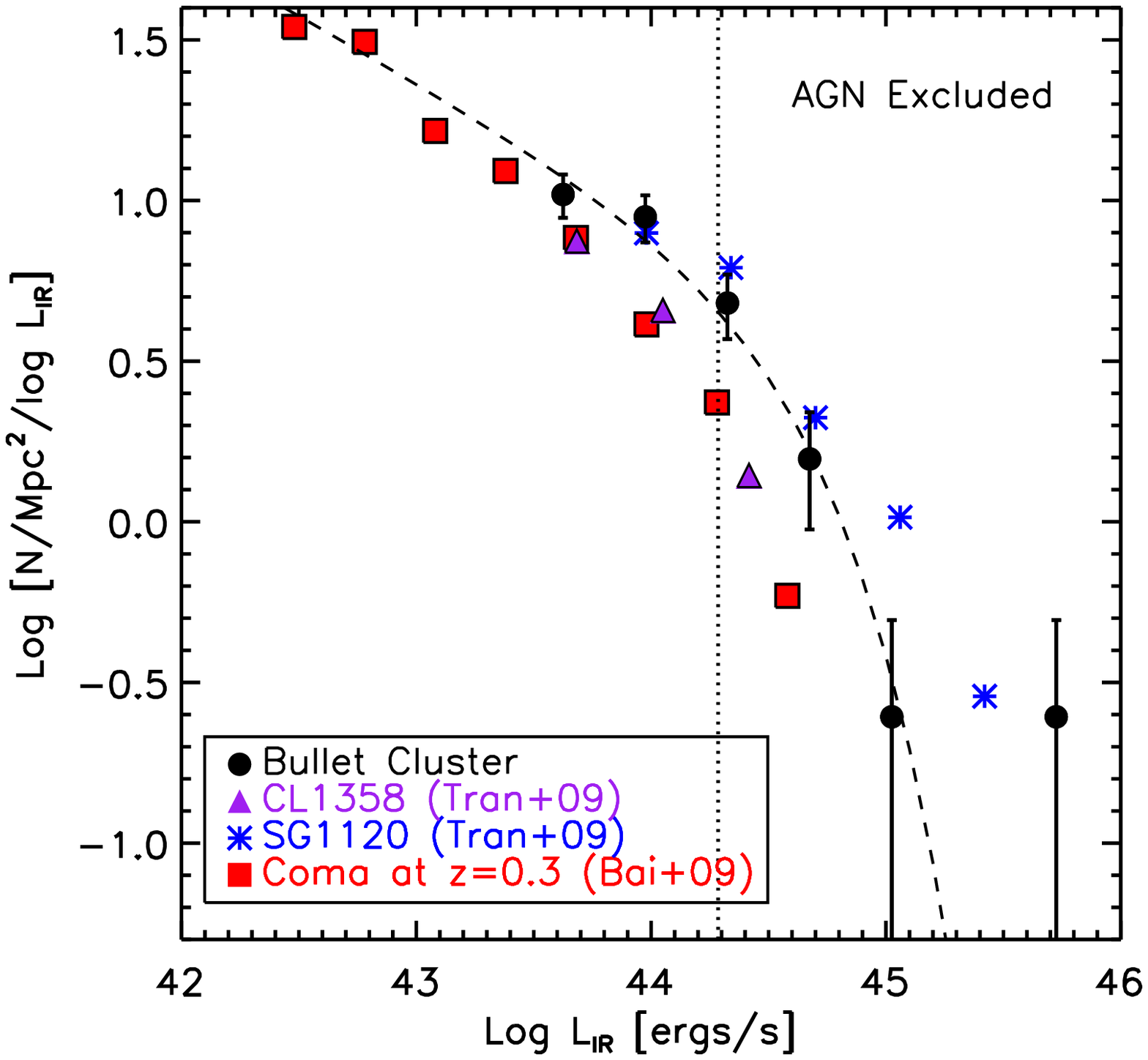}{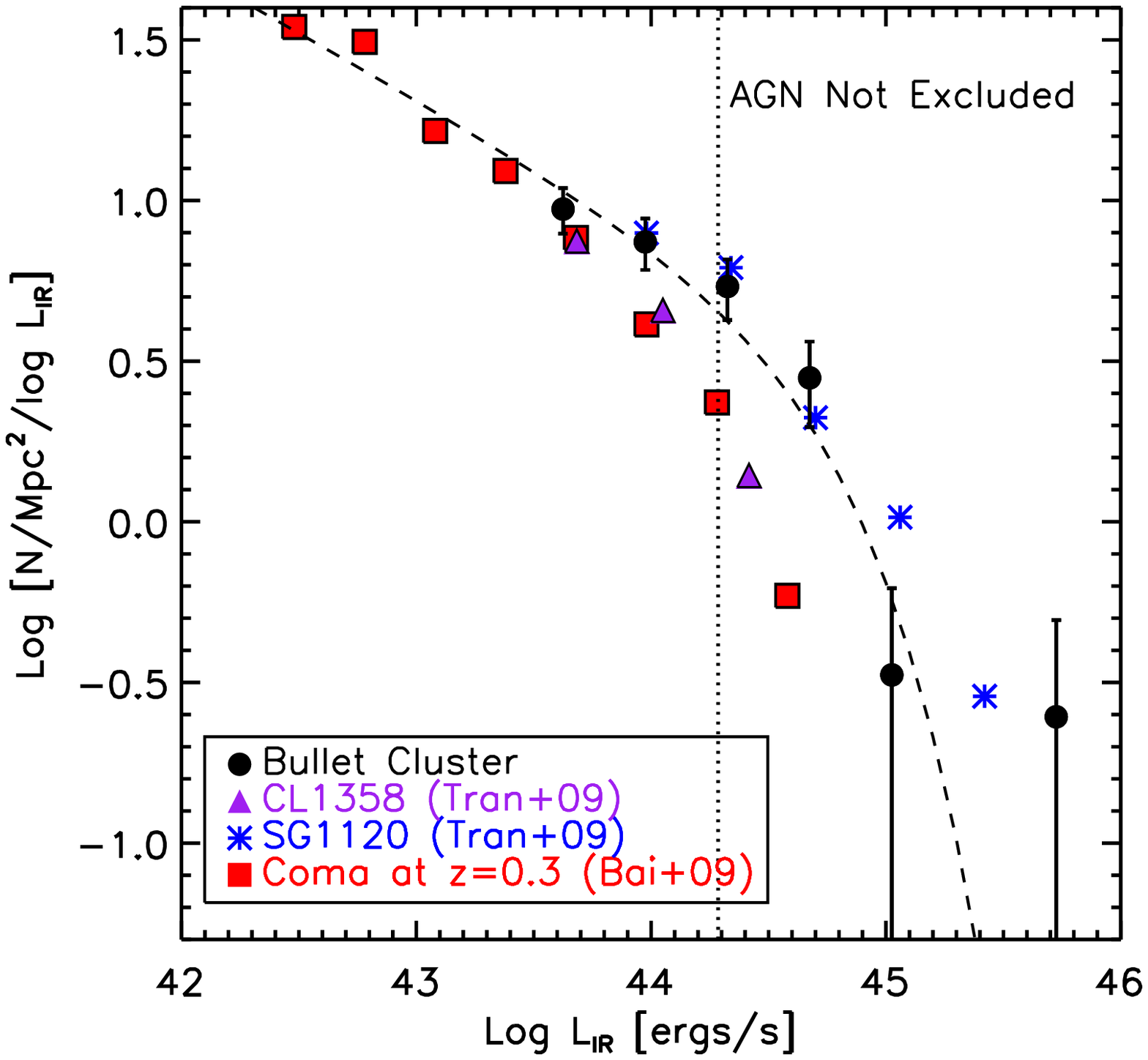}
\caption{The infrared luminosity function of the Bullet Cluster (solid circle), Coma cluster (solid square), CL1358+62 (solid triangle), and SG1120 (asterisk), with the IR LF of the Coma cluster evolved to $z\sim0.3$.  All infrared luminosities shown here are based on the \citet{dale2002} galaxy templates. The Schechter fit to the Bullet Cluster IR LF is shown as the dashed curve, and the MIPS 80\% completeness limit is indicated as a dotted vertical line.  The Left panel shows the Bullet Cluster IR LF excluding all known AGN candidates and the right panel includes six AGN confirmed members and two AGN candidate members (scaled by 0.35 to account for probability of being actual cluster members).  The inclusion of just a small number of AGN can significantly elevate the IR LF, preferentially in the bright end.  However, we note that even without the AGN, the Bullet Cluster IR LF exhibits an excess of IR luminous sources relative to Coma and CL1358+62.  The brightest Bullet Cluster galaxy is a ULIRG and is not included in the Schechter fit.}
\label{fig:IRLFv1}
\end{figure*}

The infrared luminosity function of the Bullet Cluster is presented in Figure~\ref{fig:IRLFv1}, with 1$\sigma$ Poisson error bars.  We include two versions of the Bullet Cluster IR LF -- one which excludes all known AGN wedge and X-ray point sources (left panel), and one which does not (right panel).  The aim of the latter IR LF is to illustrate the impact of AGN contamination. The first sample contains 39 confirmed cluster members and 12 cluster candidates, while the latter sample has 44 confirmed members and 14 cluster candidates.  In both cases, we weight the cluster candidates by 0.35, which is the probability that a MIPS source within the color boundaries set in Figure~\ref{fig:select} is an actual cluster member.  We apply a completeness correction derived from artificial star tests for data fainter than the MIPS 80\% completeness limit, shown as a dotted vertical line in Figure~\ref{fig:IRLFv1}.

Overplotted on Figure~\ref{fig:IRLFv1} are the IR LFs of CL1358+62 ($z=0.328$) and the Coma cluster ($z=0.024$) from \citet{tran2009} and \citet{bai2009}.  Both CL1358+62 and Coma have approximately the same total mass as the Bullet Cluster \cite[M$\sim10^{15}$\Msun;][]{kubo2007,hoekstra1998}.  The Coma IR LF shown in Figure~\ref{fig:IRLFv1} is evolved to  $z=0.3$ using the relations $L^{*}_{IR}\propto (1+z)^{3.2}$ and $\Phi^{*}_{IR}\propto (1+z)^{1.7}$ from \citet{bai2009}.  In addition to the CL1358+62 and Coma, we also overplot the IR LF of SG1120 ($z=0.37$) from \citet{tran2009}.  SG1120 consists of four galaxy groups in the process of cluster assembly, predicted to form a Coma-like cluster by $z=0$ \citep{gonzalez2005}.

The infrared luminosity function of Coma and CL1358+62 are derived from 24\micron\ MIPS data that cover  $\sim 3\times 3$ and $\sim 2.5 \times 2.5$ Mpc, respectively.  This is comparable to the Bullet Cluster MIPS data, which covers an area of $\sim 3.4\times 3.4$ Mpc.  The spatial coverage of SG1120 is $\sim6\times 6$ Mpc because the four galaxy groups require a more extensive survey area than the clusters.  In addition to the similar spatial coverage of CL1358+62, Coma, and the Bullet Cluster, we emphasize that the total mass of all systems shown in Figure~\ref{fig:IRLFv1} are comparable to the mass of the Bullet Cluster.  \citet{tran2009} do not take AGN contamination into account in the IR LFs of CL1358+62 and SG1120, while \citet{bai2009} cross-matched their Coma data with the {\it Catalogue of Quasars and Active Galactic Nuclei} \citep{veron2003}.

We fit a Schechter function to the Bullet Cluster data, fixing the faint end slope to $\alpha=-1.4$, adopted from the IR LF of Coma \citep{bai2009}.  The ULIRG, which occupies the brightest bin in Figure~\ref{fig:IRLFv1}, is excluded from the fit because it is a clear outlier from the smooth distribution of IR star-forming galaxies.  

Results from fitting a Schechter function to the Bullet Cluster data with and without AGN are summarized in Table~\ref{table:IRLF}, along with the Schechter parameters for Coma, CL1358+62 and SG1120.  We note that the Schechter parameters for the Bullet Cluster remain consistent within 1$\sigma$ whether or not we allow $\alpha$ to vary during the fit.

Figure~\ref{fig:IRLFv1} shows that the Bullet Cluster luminosity function is enhanced relative to the IR LFs of CL1358+62 and Coma at all values of $L_{IR}$.  In addition to showing a higher surface density of IR star-forming galaxies in all $L_{IR}$ bins, the IR LF of the Bullet Cluster extends to $\log L_{IR}\gap 45$ (not including the ULIRG), whereas the IR LFs of CL1358+62 and Coma truncate at $\log L_{IR}\sim 44.4$ and $\log L_{IR}\sim 44.6$, respectively.  For the IR LF excluding all known AGN, we obtain $L^{*}=44.68\pm0.11$,, which is $\sim$0.25 and 0.35 dex brighter than the  $L^{*}$ value of evolved Coma and CL1358+62, respectively.    

Figure~\ref{fig:IRLFv1} also illustrates that inclusion of even a small number of AGN sources can have a noticeable impact on the IR LF, particularly at the bright end.  If we include all known AGN, (six confirmed members and two cluster candidates weighted by 0.35 -- the probability that they are cluster members), $L^{*}$ increases to $44.81\pm0.13$, and is  $\sim$0.4 and 0.5 dex brighter than $L^{*}$ of Coma and CL1358+62, respectively.    

Although the Bullet Cluster IR LF has a value of $L^{*}$ that is brighter than that of evolved Coma or CL1358+62, it is still fainter than $L^{*}$ of SG1120 by 0.30 (0.18) dex, excluding (including) the AGN population in the Bullet Cluster sample.  It has been shown by \citet{tran2009} that SG1120 exhibits an excess of 24\micron\ sources compared to CL1358+62 and evolved $z\sim 0$ clusters, indicating that the infrared galaxies of SG1120 represent a galaxy group population whose star formation has not been quenched to typical cluster levels.  We see from Figure~\ref{fig:IRLFv1} that while the Bullet Cluster has a higher number density of IR star-forming galaxies compared to other clusters, it is still suppressed relative to these galaxy group IR LFs.

One explanation for the excess of IR bright sources in the Bullet Cluster relative to other clusters can be the presence of galaxies belonging to the infalling group population associated with the bullet, which may contribute significantly to the overall elevated IR LF.  In this case, the IR LF would be representative of a combined galaxy group and galaxy cluster environment.  To test this scenario, we scale down the SG1120 IR LF by the subcluster to SG1120 mass ratio, thereby creating a rough proxy for a galaxy group IR LF at subcluster mass.

\begin{figure}
\epsscale{1}
\plotone{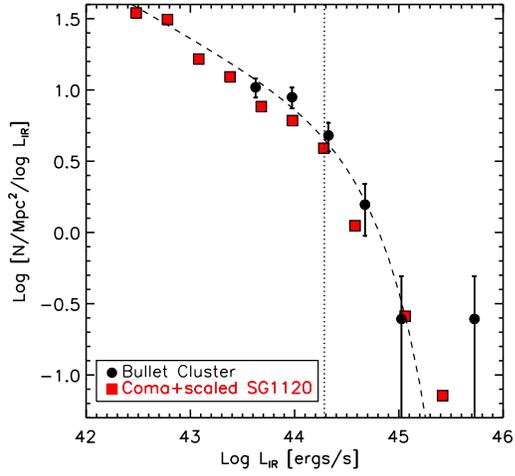}
\caption{The infrared luminosity function of the Bullet Cluster (solid circle) excluding known AGN.  Overplotted with solid (red) square is the IR LF of Coma evolved to $z=0.3$ added with the IR LF of SG1120, after the latter has been scaled down by a factor of $\sim$4 to match the approximate mass of the infalling group population in the Bullet Cluster.}
\label{fig:IRLFv2}
\end{figure}

To determine the subcluster mass, we assume a main cluster to subcluster mass ratio of 1:10 \citep{springel2007} and a total cluster mass of  $M_{200}=1.5\times 10^{15}$ (Gonzalez et al. in prep).  This yields a subcluster mass of $M_{sub}=1.4\times 10^{14}$ \Msun.  With a lower limit on the mass of SG1120 \cite[$5.3\times 10^{14}$ \Msun;][]{gonzalez2005}, the subcluster to SG1120 mass ratio is approximately 1:4. Figure~\ref{fig:IRLFv2} shows the IR LF of the Bullet Cluster overplotted with the IR LF of evolved Coma added with the IR LF of SG1120 after it has been scaled down by a factor of four.  Overall, the IR LF of combined Coma and scaled SG1120 is a good match to the observed Bullet Cluster IR LF, with the two LFs having a similar effective $L^{*}$.  This supports the idea that the Bullet Cluster IR LF is a combined distribution, representing galaxies from both a cluster and group population.

Quantitatively, the Bullet Cluster LF is still a factor of $\sim$1.5 higher than the evolved Coma+scaled SG1120 model at most luminosity bins. There are several posssible explanations. First, the intrinsic variation in the IR LF between groups remains poorly constrained--one would only need a factor of $\sim$1.5 higher surface density in the subcluster to reproduce the observed Bullet Cluster IR LF.  Second, the excess may be evidence of Êtriggered star formation in the Bullet Cluster due to the merger. ÊIn this case the factor of 1.5 can be considered an upper bound on the total enhancement induced by the merger. \citet{chung2009} found that ram pressure associated with the supersonic shock front does not have a significant impact on recent specific star formation rates of galaxies within the central R$\sim0.5$ Mpc region.  However, this does not exclude the possibility that galaxies in the cluster outskirts may experience low level triggering of star formation due to ram pressure effects upon first entering the cluster merger environment.  Since ram pressure is a fast-acting mechanism that can lead to quenched star formation in as short a time as $\sim$100 Myr, galaxies with triggered star formation in the outskirts region could already be transformed into quiescent galaxies with little to no star formation, by the time they reach the cluster core.

In either case, the overall enhancement of star-formation in the Bullet Cluster relative to quiescent clusters of similar mass appears to be dominated by the population associated with the infalling galaxy group.  This is supported by the two IR LFs shown in Figure~\ref{fig:IRLFv2}, which are consistent within a factor of 1.5.  Group to group variation in IR LFs and/or a small effect from the cluster merger itself may also contribute a small excess of IR star-forming galaxies in the Bullet Cluster.

\subsection{Lessons Learned}
We have seen throughout the previous sections that although the integrated specific SFR can be a good measure of the overall level of activity in a cluster, we must exercise caution when comparing this quantity among different clusters.  As described in \S\ref{sec:globalSFR}, the contamination from an unknown AGN population can have a significant impact on the inferred integrated SFR and the bright end of the IR LF.  Although the fraction of expected AGN in a cluster may be as low as a few percent, even a small number of AGN can dramatically increase the measured global SFR of a cluster, since AGN are preferentially IR bright compared to the normal star-forming population.  We also warn against excluding red sequence galaxies in an attempt to remove AGN from a mid-IR sample of galaxies. In the Bullet Cluster, we have found that 20\% of the non-AGN MIPS members lie on the red sequence.

In addition, it is important when interpreting the integrated specific SFRs of clusters that one has a good understanding of the infalling galaxy population.  In the case of the Bullet Cluster, a few galaxies most likely belonging to an infalling group heavily drive the integrated specific SFR, making the Bullet Cluster one of the more ``active'' clusters in Figure~\ref{fig:SFR}.  We have been able to identify the important effect of the infalling population by comparing the IR LF of the Bullet Cluster to other systems, and knowing the mass ratio of the infalling group to main cluster.  In cases where the cluster dynamics are not well characterized, particularly in cluster mergers where infalling groups may be prevalent, it is important to consider that the integrated specific SFR is not necessarily reflective of effects from the merger itself, but may be dominated by an infalling group population.


\subsection{Transformation Mechanisms}
\subsubsection{Timescale constraint from IR LF}

The IR LF of the Bullet Cluster reveals an excess of IR star-forming galaxies relative to other massive clusters.  However, the observed Bullet Cluster IR LF is consistent with a combined cluster and group IR LF (appropriately scaled for mass) within a factor of $\sim$1.5.  This supports the hypothesis that galaxies from the infalling group population (or ``subcluster'') are predominantly responsible for the enhanced star formation observed in the Bullet Cluster relative to other clusters.  In this case, we can place a lower limit on the timescale required for a cluster mechanism to quench star formation in recently acquired group galaxies.  Since subcluster star-forming galaxies are observed $\sim$250 Myr after core passage, we can rule out fast-acting mechanisms such as ram pressure, as being a dominant physical process in driving the evolution of star formation rate in a cluster merger.  Instead we require a slow-working process that can explain the excess of star-forming LIRGs observed in the Bullet Cluster 250 Myr after a major merger. 

The need for a cluster process that works to quench star formation over long timescales of a few Gyr has been supported by much observational and theoretical work.  There are two main physical mechanisms that work over such timescales in the cluster environment -- galaxy harassment and strangulation (also known as starvation).  Of these two, we prefer strangulation to be the more plausible mechanism to eventually quench the remaining star formation in the Bullet Cluster and truncate the bright end of the IR LF. Strangulation is a slow process that begins to work in the cluster outskirts.  The interaction between the cluster ICM and the loosely bound hot halo gas of a galaxy can cause the halo to be gently pushed out, eventually leading to transformations in both morphology and star formation \citep{larson1980, balogh2000}.  Simulations by \citet{bekki2002} have shown that ram pressure and global tidal effects can be an effective means of removing halo gas within a few billion years.  Once this halo gas is removed, the galaxy no longer has a source of replenishment for cold gas to continue star formation. 
  
There is much observational evidence in support of strangulation as an important cluster mechanism to explain the dearth of late type star-forming galaxies in cluster cores.  This includes work by \citet{treu2003} who classified morphologies of over 2000 galaxies in Cl0024+16 out to a 10 Mpc diameter region, and discovered ``mild gradients'' in the morphological fractions at large radii.  They argue that such a trend could only be explained by a slow working mechanism such as strangulation.  \citet{moran2006,moran2007} also support strangulation as an important cluster process, based on the spatial distribution of quenched spiral galaxies and using ultraviolet and spectroscopic signatures of star formation to show that star formation in passive spirals must have decayed over $\lap$1 Gyr.

In addition to the observational studies, there is also mounting theoretical evidence in favor of strangulation, such as work by \citet{balogh2000} who are able to reproduce the systematic observed difference in colors of cluster versus field galaxies, if they assume a gradual decline of SFR over a few Gyr in the galaxies accreted into the cluster.  It has also been shown by simulations from \citet{kawata2008} that strangulation can occur even in low mass groups, where the ICM-galaxy halo interaction is weaker than in clusters.  More recently, \citet{mcgee2009} have demonstrated through semi-analytic galaxy formation models that strangulation is a favorable mechanism to quench star formation over long timescales in order to match the observed evolution of red galaxy fractions in clusters.  Further theoretical evidence in favor of strangulation includes work by \citet{weinmann2010} who have demonstrated that models including slow removal of hot halo gas in a galaxy can match observed fractions of passive cluster satellite and central galaxies.  

Galaxy harassment is also a mechanism that works on long timescales of a few Gyr from the cumulative effect of high-speed near encounters between galaxies, which can disturb galaxy morphologies and gas content \citep{moore1996, haynes2007}.  It is a process most efficient in cluster cores, where galaxy densities are high and near encounters occur frequently.  However, all of our LIRGs lie beyond $R\sim1$ Mpc.  Assuming that the LIRGs take $\sim$1 Gyr to reach the cluster core where harassment can begin to have an impact, it is likely that other ICM processes effective in the intermediate/outskirts region (such as strangulation or galaxy-galaxy interactions) would have already begun to take effect.  In addition, it has been suggested that harassment is only mildly effective in the case of dwarf cluster galaxies ($M_{tot}\sim10^{10}$ \Msun), with only 10\% stellar mass loss on average  \citep{smith2010}.  Given that harassment has such a negligible effect on dwarf cluster galaxies, it is unlikely that it can significantly quench star formation in intermediate mass LIRGs.  

\subsubsection{Spatial Distribution of LIRGs}

In addition to using timescale constraints based on the comparison of the Bullet Cluster IR LF to other systems, we can also look at the spatial distribution of the IR bright galaxies in the Bullet Cluster to constrain mechanisms that may play a key role in triggering or quenching star formation in cluster mergers. 
Although the Bullet Cluster shows a higher number density of IR galaxies compared to other clusters at all $L_{IR}$ values in Figure~\ref{fig:IRLFv1}, it is particularly useful to examine the spatial distribution of the LIRGs/ULIRG.  Looking at the spatial distribution of these LIRGs/ULIRG is a good way to isolate the galaxies that most likely do not belong to a ``normal'' galaxy cluster IR LF.  This is evident in Figure~\ref{fig:IRLFv1}, which shows that the IR LFs of evolved Coma and CL1358+62 truncate at a much lower $L_{IR}$ than the Bullet Cluster IR LF.

\begin{figure}
\epsscale{1}
\plotone{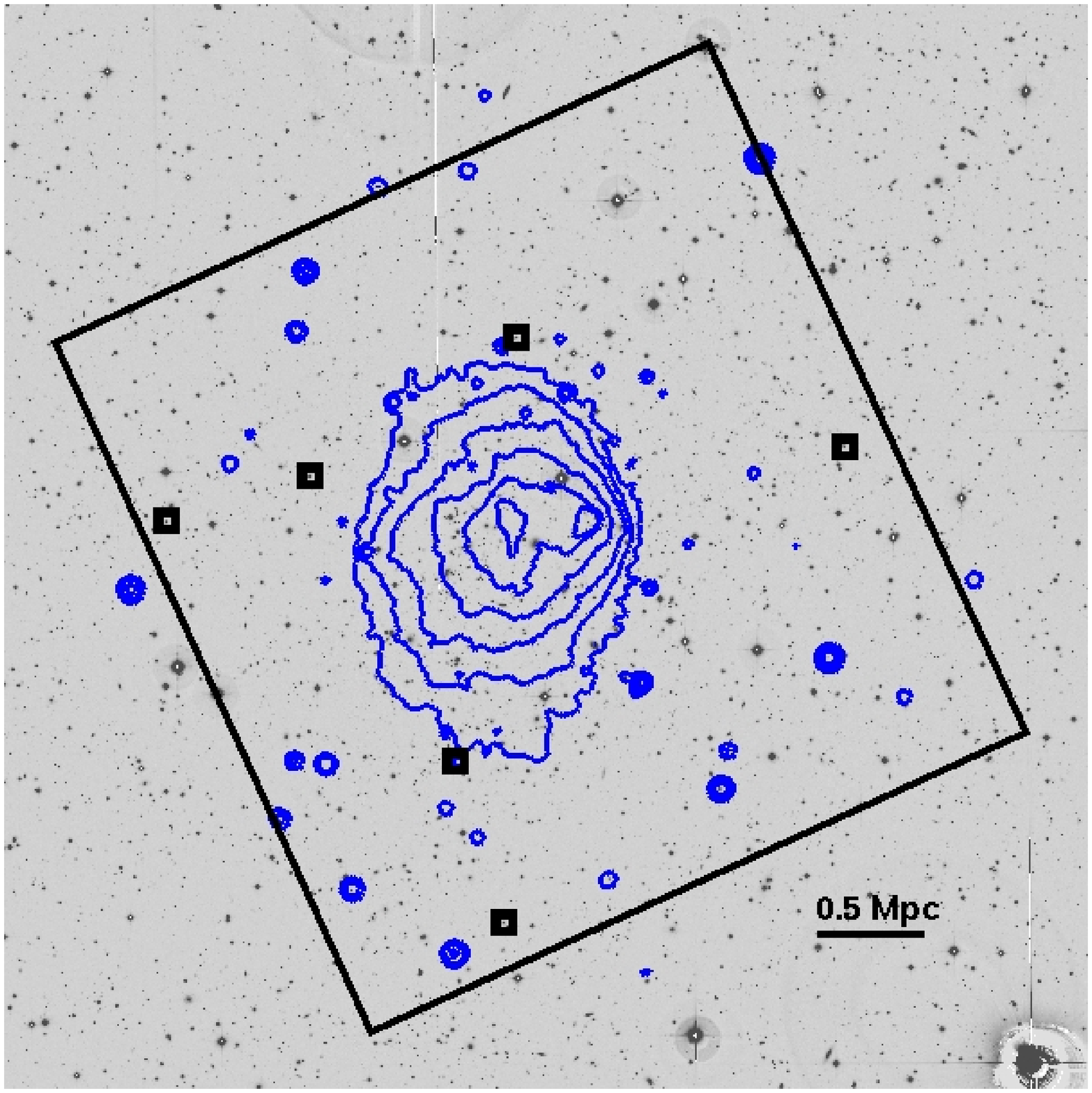}
\caption{R-band WFI image shown with X-ray surface brightness contours.  The large box shows the FOV of our 24\micron\ MIPS data, and small boxes indicate the location of the five LIRGs and one ULIRG.}
\label{fig:spatialdist}
\end{figure}

The spatial distribution of the LIRGs and ULIRG, illustrated in Figure~\ref{fig:spatialdist}, shows that all six IR bright galaxies lie beyond the central Mpc region, where a Mach 3 shock front propagates through the main cluster galaxies, creating a strong ram pressure environment. There is also no clear correlation between the distribution of the LIRGs and the cluster merger structure as shown by X-ray surface brightness contours.  This implies that the LIRGs were not triggered by the merger event itself, and in particular not triggered by ram pressure.  This is consistent with results from \citet{chung2009} who analyzed the specific SFRs of post-shock versus pre-shock galaxies in the Bullet Cluster and found that ram pressure did not significantly trigger or quench recent star formation in the Bullet Cluster galaxies.  \citet{haines2009b} similarly concluded that cluster mergers do not have a strong impact on star-forming IR galaxies by finding no correlation between the dynamical state of a cluster (whether it is relaxed or merging), and the fraction of IR luminous galaxies in a sample of 30 clusters.  In contrast, \citet{moran2005} found evidence in favor of triggered star formation traced by [OII] emission in faint early type galaxies residing near the virial radius of CL0024+16, where cluster merger effects may have generated shocks and/or induced galaxy harassment.  While \citet{moran2005} favor a fast mechanism such as merger-induced ram pressure to trigger star formation in galaxies near the cluster outskirts, they also require a gradual fading of star formation on timescales of $\sim$1 Gyr based on the radial distribution of Balmer absorption line galaxies in CL0024+16.  This is similar to the long timescale required to quench star formation in the Bullet Cluster LIRGs, if they are indeed from the subcluster population.

Galaxy-galaxy mergers or tidal interactions also do not explain the LIRG population in the Bullet Cluster, despite the preferential distribution of LIRGs in the outskirts, where the chance of mergers is relatively high within a cluster environment.  We find no obvious signs of recent interactions that would have triggered star formation from the morphologies of the LIRGs and ULIRG.   \citet{geach2009} found that in the cluster merger CL0024+16, several LIRGs belong to smaller groups that may be falling into the cluster, with some evidence that star formation is triggered by interactions within these small groups.   However, they also find no statistical difference in the local environment of LIRGs versus quiescent spiral galaxies.

The current data thus provide no evidence that the observed LIRGS are triggered by either the cluster merger or galaxy-galaxy interactions.  It also seems implausible that these LIRGS are field galaxies falling into the cluster for the first time, based on a comparison of the IR LFs of the Bullet Cluster with those of evolved Coma and CL1358+62.  The IR LFs of both evolved Coma and CL1358+62 truncate at lower values of $L_{IR}$ than in the Bullet Cluster IR LF, which would be unlikely if the Bullet Cluster LIRGs were infalling field galaxies.

Figure~\ref{fig:spatialdist} shows that the LIRGs are all distributed at $R>1$ Mpc, consistent with recent studies which find that mid-IR and sub-mm dusty star-forming galaxies avoid the cluster core and are preferentially distributed in the intermediate and outskirts regions \cite[e.g.][]{braglia2010,koyama2010}.  However, the spatial distribution of the LIRGs shown in Figure~\ref{fig:spatialdist} also corresponds to the outskirts region of the subcluster, where it is more likely to find dusty star-forming galaxies.  All the LIRGs are within 1 to 2 times the  $R_{200}$ radius of the subcluster, from the center of the ``bullet''.

Overall, the spatial distribution of the Bullet Cluster LIRGs shows no evidence of being strongly influenced by the cluster merger.  Combined with results from comparing the IR LF of the Bullet Cluster to the IR LFs of other systems, the LIRGs are unlikely to be field galaxies.  However, the spatial distribution of the LIRGs is consistent with the galaxies belonging to the outskirts region of the subcluster.

\section{Conclusions}

We use 24\micron\ MIPS data to calculate the mass normalized global star formation rate of 37 spectroscopically confirmed Bullet Cluster members within $R<1.7$ Mpc, excluding AGN sources using IRAC, X-ray, and optical spectroscopic data.  The integrated obscured SFR of the Bullet Cluster is 267 \Msolaryr\ with a 0.2 dex uncertainty, using the $L_{24}$-SFR relation from \citet{rieke2009}.  Normalizing by cluster mass we obtain a specific star formation rate of 28 \Msolaryr per $10^{14}$\Msun.  Using the \citet{dale2002} models with the \citet{kennicutt1998} SFR calibration, and applying a lower limit on 24\micron\ flux comparable to previous works, we obtain a mass normalized integrated SFR of 47 \Msolaryr per $10^{14}$\Msun.  We find that the specific SFR of the Bullet Cluster is one of the highest among other merging and non-merging clusters from the literature, with the exception of CL0024+16.  In the Bullet Cluster, a small population of LIRGs and ULIRG contribute 30\% and 40\% of the total obscured star formation rate.  These galaxies comprise the bright end of the infrared luminosity function, which are not observed in the IR LFs of evolved Coma or CL1358+62.

A Schechter fit to the IR LF of the Bullet Cluster yields $L^{*}=44.70\pm0.16$, which is $\sim$0.25 dex brighter than $L^{*}$ of evolved Coma and 0.35 dex brighter than CL1358+62, a massive cluster at $z=0.3$.  We observe an excess of IR bright sources in the Bullet Cluster along the entire range of infrared luminosities in the IR LF, compared to the IR LFs of evolved Coma and CL1358+62.  One way to explain this is to attribute the excess of IR sources to the galaxy population associated with the group that has recently been accreted into the main cluster $\sim$250 Myr ago (core passage).   The combined IR LF of evolved Coma and mass-scaled SG1120 is consistent within a factor of $\sim$1.5 of the observed Bullet Cluster IR LF.  This supports the idea that the galaxies originating from the group population have not yet been processed to match typical cluster levels of star formation rate.  In this case, a transformation mechanism such as strangulation, which acts on timescales longer than the merger timescale of the main cluster and subcluster ($\gap250$ Myr) is necessary to eventually truncate the bright end and suppress the faint end of the IR LF to match the observed distribution in Coma and CL1358+62.


\begin{deluxetable}{cccc}
\tabletypesize{\scriptsize}
\tablecaption{Data}
\tablewidth{0pt}
\tablehead{
\colhead{Name} & \colhead{MIPS} & \colhead{MIPS AGN} & \colhead{ALL}
}
\startdata
 Confirmed members  &  44  &  6  &  406 \\
 Interlopers & 83   &  5 &  495 \\
 Candidate members & 14 & 2 &  ...\\
\enddata
\label{table:data}
\tablecomments{The number of galaxies that are spectroscopicaly confirmed cluster members, interlopers and cluster candidates (based on optical and IRAC colors). The columns show number of galaxies from the MIPS catalog with WFI and IRAC counterparts, number of AGN within the MIPS catalog determined by IRAC colors, X-ray emission, and optical spectroscopy, and the total number of galaxies from the entire IMACS spectroscopic catalog.} 

\end{deluxetable}

\begin{deluxetable*}{ccccc}
\tabletypesize{\scriptsize}
\tablecaption{LIRGs and AGN}
\tablewidth{0pt}
\tablehead{
\colhead{Object Type} & \colhead{RA [hh:mm:ss]} & \colhead{Dec [$^{\circ}\ \arcmin\ \arcsec$]} & \colhead{$L_{IR}$ [\Lsun]} &  \colhead{SFR [\Msolaryr] }
}
\startdata
ULIRG/LINER & 6:58:30.87  &  -56:03:36.31 & $1.3\times 10^{12}$   &   104  \\ 
LIRG & 6:59:13.45  &  -55:56:31.19 & $2.5\times 10^{11}$   &   28 \\
LIRG & 6:58:55.33  &  -55:55:43.04 & $1.5\times 10^{11}$   &   16  \\
LIRG & 6:58:29.30  &  -55:53:15.92 & $1.4\times 10^{11}$   &   14  \\ 
LIRG & 6:57:47.86  &  -55:55:10.83 & $1.1\times 10^{11}$   &   11  \\ 
LIRG &  6:58:37.06  &  -56:00:45.66  &  $1.0\times 10^{11}$  & 10 \\ 
LIRG, X-ray AGN& 6:58:03.52  &  -56:01:14.23 &  $1.0\times 10^{11}$  & ... \\ 
X-ray \& IRAC AGN &  6:58:35.22  &  -56:01:4.78  & $8.7\times 10^{10}$  & ... \\ 
X-ray AGN &  6:58:17.48  &  -56:02:49.35  &  $8.1\times 10^{10}$  &  ... \\ 
Optical AGN (Seyfert) &  6:58:42.50  &  -56:00:28.22 & $7.6\times 10^{10}$ &  ... \\
IRAC AGN &  6:58:08.48  &  -55:53:36.45 &  $1.7\times 10^{10}$  &  ...  \\
X-ray AGN & 6:58:26.67  &  -56:00:0.18  &  $9.2\times 10^{9}$ &  ... \\

\enddata
\label{table:LIRGs}
\tablecomments{Spectroscopically confirmed members with MIPS emission that are either LIRGs and/or AGN.  Column 1 notes whether a galaxy is a LIRG/ULIRG, and if it is detected as either an X-ray point source or an IRAC AGN wedge source.  Columns 2-5 show the right ascension, declination, total infrared luminosity, and star formation rate for each galaxy.  SFR of the ULIRG assumes that 35\% of the total IR luminosity is powered by AGN, and the rest by star formation.  For the LIRGs and AGN, the 1$\sigma$ error of the measured 24\micron\ flux  is between 1-10\%, and thus the final error of the total infrared luminosity is dominated by the uncertainty in the $L_{24}$ to $L_{IR}$ calibration, which is $\lap$0.15 dex.}
\end{deluxetable*}


\begin{deluxetable}{cccc}
\tabletypesize{\scriptsize}
\tablecaption{Schechter Parameters}
\tablewidth{0pt}
\tablehead{
\colhead{Cluster/group} & \colhead{$\alpha$} & \colhead{$\log L^{*}$} & \colhead{$\phi^{*}$ [N/Mpc$^{2}$/log $L_{IR}$]}
}
\startdata
Coma (evolved $z=0.3)$   &  -1.4  &  $44.44^{+0.27}_{-0.24}$  &  ... \\
CL1358+62 & -1.4  & $44.33^{+0.32}_{-0.25}$ &  $1.92^{+1.4}_{-1.3}$ \\
SG1120 & -1.4 & $44.99^{+0.19}_{-0.19}$ &  $1.25^{+0.5}_{-0.3}$   \\
Bullet Cluster & -1.4 & $44.68^{+0.11}_{-0.11}$ & $1.77^{+0.5}_{-0.5}$ \\  
Bullet Cluster (with AGN) & -1.4 & $44.81^{+0.13}_{-0.13}$ &  $1.36^{+0.4}_{-0.4}$ \\ 
\enddata
\label{table:IRLF}
\tablecomments{Schechter parameters for the clusters and SG1120 shown in Figure~\ref{fig:IRLFv1}.  Schechter parameters for Coma are adopted from \citet{bai2009} and evolved to $z=0.3$ using $L^{*}\propto(1+z)^{3.2}$ evolution, while values for CL1358+62 and SG1120 are from \citet{tran2009}.  We fit a Schechter function to the observed Bullet Cluster IR LF, without and with the AGN, fixing the faint end slope $\alpha$ to the match that of Coma.}
\end{deluxetable}


\acknowledgements This work is based on observations made with the Spitzer Space
Telescope, which is operated by the Jet Propulsion Laboratory,
California Institute of Technology under a contract with NASA. Support
for this work was provided by NASA through an award issued by
JPL/Caltech.  The authors acknowledge support for this work from NASA/Spitzer grants 1319141 and 1376614.

{\it Facilities:} \facility{Spitzer (IRAC)}, \facility{Spitzer (MIPS)}, \facility{CXO (ACIS-I)}, \facility{Magellan:Baade (IMACS)}

\end{document}